\documentclass[12pt]{article}
\usepackage{amssymb,amsmath,epsfig}
\allowdisplaybreaks

\begin{document}
\title{\bf Gravitational Decoupled Charged Anisotropic Spherical Solutions}
\author{M. Sharif \thanks{msharif.math@pu.edu.pk} and Sobia Sadiq
\thanks{sobiasadiq.01@gmail.com}\\
Department of Mathematics, University of the Punjab,\\
Quaid-e-Azam Campus, Lahore-54590, Pakistan.}

\date{}
\maketitle
\begin{abstract}
The purpose of this paper is to obtain exact solutions for charged
anisotropic spherically symmetric matter configuration. For this
purpose, we consider known solution for isotropic spherical system
in the presence of electromagnetic field and extend it to two types
of anisotropic charged solutions through gravitational decoupling
approach. We examine physical characteristics of the resulting
models. It is found that only first solution is physically
acceptable as it meets all the energy bounds as well as stability
criterion. We conclude that stability of the first model is enhanced
with the increase of charge.
\end{abstract}
{\bf Keywords:} Exact Solutions; Anisotropy; Gravitational
Decoupling.\\
{\bf PACS:} 04.20.Jb; 04.40.-b; 04.40.Nr.

\section{Introduction}

General relativity is one of the cornerstones that provides basic
understanding of astrophysical phenomena as well as the cosmos. The
structure of self-gravitating systems is attained by solving the
famous Einstein field equations. Schwarzschild was the first who
determined vacuum solution of these equations describing the
geometry of region exterior to a prefect fluid sphere in hydrostatic
equilibrium. Tolman \cite{2} found several solutions by solving the
field equations for static sphere of perfect fluid with cosmological
constant and discussed the matching of resulting interior solutions
with exterior one. After that, many exact solutions for isotropic
and anisotropic static as well as non-static configurations have
been obtained \cite{3}.

The formulation of interior solutions for self-gravitating systems
is a difficult task due to the existence of non-linearity in the
field equations. In this regard, the minimal gravitational
decoupling (MGD) approach has been very useful in finding exact and
physically feasible solutions for spherically symmetric stellar
distributions. This strategy was genuinely proposed by Ovalle
\cite{4} to find an exact solution for compact stars in the context
of the braneworld. In this framework, Ovalle and Linares \cite{5}
developed an exact interior solution to the field equations for
isotropic spherically symmetric compact distribution. They concluded
that this solution represents braneworld version of the Tolman IV
solution \cite{2}. Casadio et al. \cite{6} used MGD concept by
modifying the temporal as well as radial metric function and found a
new exterior solution for spherical self-gravitating system which
presents a naked singularity at the Schwarzschild radius. Ovalle
\cite{7} decoupled gravitational sources to construct anisotropic
solutions from perfect fluid solutions with spherical symmetry.
Ovalle et al. \cite{8} extended isotropic interior solution \cite{2}
by means of MGD for static stellar models to include the effects of
anisotropy.

In astrophysical context, pressure anisotropy (generated by various
physical phenomena) plays a key role in the evolution of stellar
bodies. Mak and his collaborators \cite{9} obtained exact solutions
by taking a particular form of anisotropy (difference of radial and
tangential pressures) and found that spherical star supports
positive and finite density as well as pressures. They also
discussed that the obtained radius and mass can describe realistic
astrophysical objects. Gleiser and Dev \cite{10} explored the
existence of anisotropic self-gravitating sphere and found that
anisotropy can support stars with compactness $2M/R=8/9$ ($M$ and
$R$ represent mass and radius, respectively). They also concluded
that stable configurations exist for smaller values of the adiabatic
index as compared to isotropic fluid. Sharma and Maharaj \cite{11}
obtained some exact solutions for spherically symmetric anisotropic
matter distribution satisfying linear equation of state (EoS) to
describe compact stars. We investigated the equilibrium structure of
static spherical as well as cylindrical polytropic configurations
with anisotropic source \cite{12}. Azam et al. \cite{13} generalized
these structures for generalized polytropic EoS.

The inclusion of electromagnetic field in stellar models is very
fascinating in describing their evolution. Xingxiang \cite{14}
discussed the characteristics of an exact solution for static
spherical symmetry with charged perfect fluid distribution. Di
Prisco et al. \cite{15} investigated the impact of electromagnetic
field on the dynamics of imperfect collapsing sphere and discussed a
relationship between the Weyl tensor and inhomogeneity of energy
density. Sharif and Bhatti \cite{16} analyzed the behavior of
physical parameters and energy conditions for charged anisotropic
spherically symmetric solutions. Takisa and Maharaj \cite{17}
formulated exact solutions of the Einstein-Maxwell field equations
with polytropic EoS which can be used to model charged anisotropic
compact objects. Singh and Pant \cite{18} presented charged
anisotropic spherical solution and found that the developed model is
stable and well-behaved for a wide range of anisotropy as well as
charge parameter. They also obtained that charged anisotropic
neutron and quark stars with large masses can be modeled from the
resulting solution. Khan et al. \cite{19} studied the effects of
charge on anisotropic spherical collapse with cosmological constant
and concluded that electromagnetic field enhances the rate of
destruction.

The significance of relativistic models is based on their stable
structure. Herrera \cite{20} proposed the notion of cracking as well
as overturning (when the total radial forces in a system reverse
their signs from positive to negative, cracking occurs while the
opposite situation experiences overturning) to investigate the
behavior of isotropic and anisotropic configurations just after the
equilibrium state is perturbed. He concluded that perfect fluid
distribution remains stable while cracking appears in anisotropic
case. Abreu et al. \cite{21} broadened the idea of cracking by means
of sound speed for anisotropic spherical configuration and concluded
that the system is unstable for $v_{st}^2>v_{sr}^2$ ($v_{sr}^2$ and
$v_{st}^2$ indicate radial and tangential sound speeds,
respectively). We explored the stability of charged anisotropic
polytropes and found that compact object remains stable for a
reasonable choice of perturbed polytropic index \cite{22}. Mardan
and Azam \cite{23} examined the stability of charge anisotropic
cylindrical system admitting generalized polytropic EoS and
concluded that the constructed model is unstable for several choices
of polytropic parameters.

In this paper, we explore exact charged anisotropic spherical
solutions using a known charged isotropic solution with MGD
approach. The plan of the paper is as follows. In section
\textbf{2}, we deal with the basic formalism of MGD and formulate
the corresponding field equations. The matching of interior
spacetime with the exterior one is also investigated. In section
\textbf{3}, we obtain two types of exact solutions for anisotropic
spherical source in the presence of electromagnetic field and
investigate physical characteristics of all solutions. Finally, we
conclude our results in the last section.

\section{Fluid Configuration and MGD Approach}

We consider static spherically symmetric spacetime describing the
interior geometry as
\begin{equation}\label{1}
ds^2_{-}=-e^{\eta_{-}(r)}dt^{2}+e^{\chi_{-}(r)}
dr^{2}+r^{2}\left(d\theta^{2}+\sin^{2}\theta d\phi^2\right).
\end{equation}
The energy-momentum tensor for internal constitution is given as
\begin{equation}\label{2}
T_{\alpha\beta}^{(tot)}=T_{\alpha\beta}^{(m)}+\alpha\Theta_{\alpha\beta},
\end{equation}
where
\begin{equation}\label{3}
T_{\alpha\beta}^{(m)}=(\rho+P)V_\alpha
V_\beta+Pg_{\alpha\beta}+\frac{1}{4\pi}\left(F_{\alpha}~^{\mu}F_{\beta\mu}-
\frac{1}{4}F^{\mu\nu}F_{\mu\nu}g_{\alpha\beta}\right),
\end{equation}
which represents charged perfect fluid distribution with $\rho,~P$
and $V_{\alpha}$ indicating the density, pressure and four velocity,
respectively. The term $\Theta_{\alpha\beta}$ is an additional
source coupled to gravity through constant $\alpha$ which may
contain some new fields (scalar, vector or tensor) and generate
anisotropies in self-gravitating bodies. In Eq.(\ref{3}),
$F_{\alpha\beta}=\phi_{\beta,\alpha}-\phi_{\alpha,\beta}$ is the
Maxwell field tensor and $\phi_\alpha$ is four potential. The
Maxwell field tensor satisfies the following field equations
\begin{equation*}
F^{\alpha\beta}_{~~;\beta}={\mu}_{0}J^{\alpha},\quad
F_{[\alpha\beta;\gamma]}=0,
\end{equation*}
here, $\mu_0$ is the magnetic permeability and $J^{\alpha}$ is the
four current. In comoving coordinates, we have
\begin{equation*}
\phi_{\alpha}={\phi} {\delta^{0}_{\alpha}},\quad J_{\alpha}=\zeta
V_{\alpha},\quad V^{\alpha}=e^{-\eta/2}\delta_{0}^{\alpha},
\end{equation*}
where $\zeta=\zeta(r)$ and $\phi=\phi(r)$ represent scalar potential
and charge density, respectively. The Maxwell field equation for our
spacetime yields
\begin{equation*}
\phi^{\prime\prime}+\left(\frac{2}{r}-\frac{\eta'}{2}-\frac{\chi'}{2}\right)\phi'=
4\pi\zeta e^{\frac{\eta}{2}+\chi},
\end{equation*}
where prime denotes differentiation with respect to $r$. Integration
of the above equation yields
\begin{equation*}
{\phi}'=\frac{e^{\frac{\eta+\chi}{2}}q(r)}{r^{2}}.
\end{equation*}
Here $q(r)=4\pi\int^r_{0}{\zeta}e^{\frac{\chi}{2}}{r^2}dr$ indicates
total charge inside the sphere.

The Einstein-Maxwell field equations corresponding to Eqs.(\ref{1})
and (\ref{2}) turn out to be
\begin{eqnarray}\label{4}
&&8\pi\left(\rho+\frac{q^2}{8\pi r^4}-\alpha\Theta_{0}^{0}\right)
=\frac{1}{r^2}+e^{-\chi}\left(\frac{\chi'}{r}-\frac{1}{r^{2}}\right),\\\label{5}
&&8\pi\left(P-\frac{q^2}{8\pi r^4}+\alpha\Theta_{1}^{1}\right)
=-\frac{1}{r^2}+e^{-\chi}\left(\frac{\eta'}{r}+\frac{1}{r^{2}}\right),\\\label{6}
&&8\pi\left(P+\frac{q^2}{8\pi
r^4}+\alpha\Theta_{2}^{2}\right)=e^{-\chi}\left(\frac{\eta''}{2}+\frac{\eta'^{2}}{4}-\frac{\eta'\chi'}{4}
+\frac{\eta'}{2r}-\frac{\chi'}{2r}\right).
\end{eqnarray}
The equilibrium structure of stellar object is described by
hydrostatic equilibrium equation obtained through the conservation
of energy-momentum tensor ($T^{(tot)\alpha}_{~~~~\beta;\alpha}=0$)
as
\begin{eqnarray}\label{7}
\frac{dP}{dr}+\alpha\frac{d\Theta_{1}^{1}}{dr}+\frac{\eta'}{2}(\rho+P)+\frac{\alpha\eta'}{2}
(\Theta_{1}^{1}-\Theta_{0}^{0})
+\frac{2\alpha}{r}(\Theta_{1}^{1}-\Theta_{2}^{2})-\frac{qq'}{4\pi
r^4}=0.
\end{eqnarray}
We see that Eqs.(\ref{4})-(\ref{7}) form a system of four non-linear
differential equations consisting of eight unknowns
($\eta,~\chi,~\rho,~P,~q,~\Theta_{0}^{0},~\Theta_{1}^{1},~\Theta_{2}^{2}$).
In order to find these unknowns, we follow a systematic scheme
developed by Ovalle \cite{8}. From Eqs.(\ref{4})-(\ref{6}), we
identify the matter components as
\begin{eqnarray}\label{8}
\bar{\rho}=\rho-\alpha\Theta_{0}^{0},\quad\bar{P}_{r}=P+\alpha\Theta_{1}^{1},
\quad\bar{P}_{t}=P+\alpha\Theta_{2}^{2},
\end{eqnarray}
where $\bar{\rho},~\bar{P}_{r},~\bar{P}_{t}$ represent effective
energy density, radial/tangential pressure, respectively. This shows
that the source $\Theta_{\alpha\beta}$ can produce anisotropy
$\bar{\Delta}=\bar{P}_{t}-\bar{P}_{r}=\alpha(\Theta_{1}^{1}-\Theta_{2}^{2})$
in the interior of stellar distribution.

Now, we consider the MGD approach to solve the system of
Eqs.(\ref{4})-(\ref{6}). The basic ingredient of MGD is to consider
a perfect fluid solution ($\xi,~\mu,~\rho,~P,~q$) for the
line-element given as
\begin{equation}\nonumber
ds^2=-e^{\xi(r)}dt^{2}+\frac{dr^{2}}{\lambda(r)}+r^{2}\left(d\theta^{2}+\sin^{2}\theta
d\phi^2\right),
\end{equation}
where $\lambda=1-\frac{2m}{r}+\frac{q^2}{r^2}$ with $m$ representing
the Misner-Sharp mass of fluid configuration. In order to
incorporate the effects of source $\Theta_{\alpha\beta}$ in charged
isotropic model, we consider the geometric deformation as \cite{8}
\begin{equation}\label{8b}
\xi\rightarrow\eta = \xi,\quad\lambda\rightarrow
e^{-\chi}=\lambda+\alpha g^{*},
\end{equation}
where $g^{*}$ is the deformation endured by radial metric function.
Making use of the above radial coefficient, the field equations can
be divided into two sets. The first set is given as
\begin{eqnarray}\label{9}
&&8\pi\rho+\frac{q^2}{r^4}
=\frac{1}{r^2}+e^{-\chi}\left(\frac{\chi'}{r}-\frac{1}{r^{2}}\right),\\\label{10}
&&8\pi P-\frac{q^2}{r^4}
=-\frac{1}{r^2}+e^{-\chi}\left(\frac{\eta'}{r}+\frac{1}{r^{2}}\right),\\\label{11}
&&8\pi
P+\frac{q^2}{r^4}=e^{-\chi}\left(\frac{\eta''}{2}+\frac{\eta'^{2}}{4}-\frac{\eta'\chi'}{4}
+\frac{\eta'}{2r}-\frac{\chi'}{2r}\right),
\end{eqnarray}
while the second one is
\begin{eqnarray}\label{12}
&&\alpha\Theta_{0}^{0}=\frac{g^{*'}}{r}+\frac{g^{*}}{r^{2}},\\\label{13}
&&\alpha\Theta_{1}^{1}=g^{*}\left(\frac{\eta'}{r}+\frac{1}{r^{2}}\right),\\\label{14}
&&\alpha\Theta_{2}^{2}=g^{*}\left(\frac{\eta''}{2}+\frac{\eta'^{2}}{4}-\frac{\eta'\chi'}{4}
+\frac{\eta'}{2r}-\frac{\chi'}{2r}\right).
\end{eqnarray}
The above set of equations looks like the field equations for
anisotropic spherical source
($\bar{\rho}=\Theta_{0}^{*0}=\Theta_{0}^{0}-\frac{1}{8\pi
r^2},~\bar{P}_{r}=\Theta_{1}^{*1}=\Theta_{1}^{1}-\frac{1}{8\pi
r^2},~\bar{P}_{t}=\Theta_{2}^{*2}=\Theta_{2}^{2}$) with the metric
\begin{equation}\nonumber
ds^2=-e^{\eta}dt^{2}+\frac{dr^{2}}{g^{*}}+r^{2}\left(d\theta^{2}+\sin^{2}\theta
d\phi^2\right).
\end{equation}

The matching of interior and exterior regions is obtained by
junction conditions which yield a smooth matching of two regions and
play a vital role in the study of evolution of relativistic objects.
If we consider the general outer metric as
\begin{eqnarray}\nonumber
ds^2_{+}=-e^{\eta_{+}}dt^{2}+e^{-\chi_{+}}dr^{2}+r^{2}\left(d\theta^{2}+\sin^{2}\theta
d\phi^2\right),
\end{eqnarray}
then the first fundamental form ($[ds^{2}]_{\Sigma}=0,~\Sigma$
\text{represents the hypersurface}) of junction conditions yield
\begin{eqnarray}\label{15}
\eta_{-}(R)=\eta_{+}(R),\quad
1-\frac{2M_{0}}{R}+\frac{Q_{0}^{2}}{R^2}+\alpha
g^{*}(R)=e^{-\chi_{+}(R)},
\end{eqnarray}
where $\lambda=e^{-\chi}-\alpha g^{*}$ has been used. Here, $M_{0}$
and $Q_{0}$ indicate total mass and charge within the sphere,
respectively. The second fundamental form
($[T_{\alpha\beta}S^{\beta}]_{\Sigma}=0$, $S^{\beta}$ is the unit
four-vector in radial direction) \cite{8} gives
\begin{eqnarray}\nonumber
P(R)-\frac{Q_{0}^{2}}{8\pi
R^4}+\alpha(\Theta_{1}^{1}(R))_{-}=\alpha(\Theta_{1}^{1}(R))_{+},
\end{eqnarray}
which leads to
\begin{equation}\label{16}
P(R)-\frac{Q_{0}^{2}}{8\pi R^4}+\frac{\alpha
g^{*}(R)}{8\pi}\left(\frac{1}{R^2}+\frac{\eta'(R)}{R}\right)=\frac{\alpha
h^{*}(R)}{8\pi R^2}
\left(1+\frac{2\mathcal{M}R-2\mathcal{Q}^2}{\left(R^2
-2\mathcal{M}R+\mathcal{Q}^2\right)}\right),
\end{equation}
where $h^{*}$ describes deformation in the radial metric function of
Riessner-Nordstr\"{o}m (RN) spacetime while $\mathcal{M}$ and
$\mathcal{Q}$ indicate mass and charge in the exterior region. The
necessary and sufficient conditions for the smooth matching of
interior MGD metric with spherically symmetric exterior described by
deformed RN line-element (which can be filled with fields contained
in source $\Theta_{\alpha\beta}$) are given by Eqs.(\ref{15}) and
(\ref{16}). If the exterior geometry is considered as the standard
RN metric, Eq.(\ref{16}) yields
\begin{eqnarray}\label{17}
\bar{P}(R)-\frac{Q_{0}^{2}}{8\pi R^4}\equiv
P(R)-\frac{Q_{0}^{2}}{8\pi R^4}+\frac{\alpha
g^{*}(R)}{8\pi}\left(\frac{1}{R^2}+\frac{\eta'(R)}{R}\right)=0.
\end{eqnarray}

In the following, we consider a known solution of isotropic
spherical system in the presence of charge to continue our
systematic analysis.

\section{Anisotropic Solutions}

A crucial ingredient in obtaining the anisotropic solutions using
MGD approach is to consider solution of the field equations for
spherically symmetric charged perfect fluid configuration. For this
purpose, we consider Krori and Barua's solution \cite{24} given as
\begin{eqnarray}\label{18}
&&e^{\eta}=e^{Br^2+C},\\\label{19}
&&e^{\chi}=\lambda^{-1}=e^{Ar^2},\\\label{20}
&&\rho=\frac{e^{-Ar^2}}{16\pi}\left(5A-B(B-A)r^2-\frac{1}{r^2}\right)
+\frac{1}{16\pi
r^2},\\\label{21}
&&P=\frac{e^{-Ar^2}}{16\pi}\left(4B-A+B(B-A)r^2+\frac{1}{r^2}\right)
-\frac{1}{16\pi
r^2},\\\label{22}
&&q^{2}=\frac{e^{-Ar^2}}{2r^4}\left(B(B-A)r^2-A-\frac{1}{r^2}\right)
+\frac{1}{2r^6},
\end{eqnarray}
where $A,~B$ and $C$ are constants that can be determined from
matching conditions. The rationale behind the choice of the above
solution lies in a fact that it is singularity-free and satisfies
physical conditions inside the sphere. For RN spacetime as an
exterior geometry, the matching conditions yield
\begin{eqnarray}\label{23}
A&=&\frac{-\ln(1-\frac{2M_{0}}{R}+\frac{Q_{0}^{2}}{R^2})}{R^2},\quad
B=\frac{2M_{0}R-Q_{0}^{2}}{2R^2(R^2+Q_{0}^{2}-2M_{0}R)},\\\label{24}
C&=&\frac{1}{2}\left\{1+2\ln\left(1-\frac{2M_{0}}{R}+\frac{Q_{0}^{2}}
{R^2}\right)-\frac{R^2}{R^2-2M_{0}R+Q_{0}^2}\right\},
\end{eqnarray}
with the compactness parameter $\frac{M_{0}}{R}<\frac{4}{9}$. The
above equations ensure continuity of the interior solution
(\ref{18})-(\ref{22}) with the exterior region at the boundary and
will definitely be changed after adding the source
$\Theta_{\alpha\beta}$ in the interior of sphere.

Now we move towards anisotropic solutions and turn $\alpha$ on in
the interior. The temporal and radial metric coefficients are given
by Eqs.(\ref{18}) and (\ref{8b}), respectively, while the
deformation $g^{*}$ is related to $\Theta_{\alpha\beta}$ through
Eqs.(\ref{12})-(\ref{14}) whose solution will be determined by
specifying an additional constraint. In order to achieve this goal,
we impose some conditions and find two exact solutions.

\subsection{Solution I}

Here, we apply a constraint on $\Theta_{1}^{1}$ and find solution of
the field equations for $g^{*}$ and $\Theta_{\alpha\beta}$. From
Eq.(\ref{17}), we see that RN exterior solution is compatible with
isotropic interior matter as long as $P(R)-\frac{Q_{0}^2}{8\pi
R^4}\sim\alpha(\Theta_{1}^{1}(R))_{-}$. Thus the simplest choice is
to take
\begin{equation}\label{24}
\Theta_{1}^{1}=P-\frac{q^2}{8\pi
r^4}~\Rightarrow~g^{*}=\lambda-\frac{1}{1+r\eta'},
\end{equation}
where Eqs.(\ref{10}) and (\ref{13}) have been used. The above
equation leads to the radial metric function as
\begin{equation}\label{25}
e^{-\chi}=(1+\alpha)\lambda-\frac{\alpha}{1+2r^2B}.
\end{equation}
The metric functions of interior spacetime in Eqs.(\ref{18}) and
(\ref{25}) represent the Krori and Barua solution minimally deformed
by the generic anisotropic source $\Theta_{\alpha\beta}$. It is
worthwhile to mention here that $\alpha\rightarrow 0$ leads to the
standard isotropic charged spherical solution
((\ref{18})-(\ref{22})).

The continuity of first fundamental form of matching conditions
yields
\begin{eqnarray}\label{26}
\ln\left(1-\frac{2\mathcal{M}}{R}+\frac{\mathcal{Q}^2}{R^2}\right)
&=&Br^2+C,\\\label{27}
1-\frac{2\mathcal{M}}{R}+\frac{\mathcal{Q}^2}{R^2}&=&(1+\alpha)\lambda
-\frac{\alpha}{1+2R^2B},
\end{eqnarray}
while the continuity of second fundamental form
($P(R)-\frac{Q_{0}^2}{8\pi R^4}+\alpha(\Theta_{1}^{1}(R))_{-}=0$)
leads to
\begin{equation}\label{28}
P(R)-\frac{Q_{0}^{2}}{8\pi R^4}=0\quad\Rightarrow\quad A=
\frac{\ln(2BR^2+1)}{R^2},
\end{equation}
where the constraint in Eq.(\ref{24}) has been used. Eliminating
$\frac{2\mathcal{M}}{R}$ from Eq.(\ref{27}), we find
\begin{equation}\label{29}
\frac{2\mathcal{M}}{R}=\frac{2M_{0}}{R}+\frac{\mathcal{Q}^{2}
-Q_{0}^{2}}{R^2}-\alpha\left(1-\frac{2M_{0}}{R}+\frac{Q_{0}^2}{R^2}\right)+\frac{\alpha}{1+2R^2B}.
\end{equation}
Inserting the above equation in (\ref{26}), we obtain
\begin{equation}\label{30}
BR^2+C=\ln\left(1-\frac{2M_{0}}{R}+\frac{Q_{0}^{2}}{R^2}
+\alpha\left(1-\frac{2M_{0}}{R}+\frac{Q_{0}^2}{R^2}\right)-\frac{\alpha}{1+2R^2B}\right),
\end{equation}
which yields the constant $C$ in terms of $B$. Here, the necessary
and sufficient conditions for the smooth matching of interior and
exterior metrics are given by Eqs.(\ref{28})-(\ref{30}). In this
case, the anisotropic solution, i.e., the expressions of
$\bar{\rho},~\bar{P}_{r},~\bar{P}_{t},~\bar{\Delta}$ and $q$ are
obtained as
\begin{eqnarray}\nonumber
\bar{\rho}&=&\frac{1+4B^2r^4+4Br^2(1-\alpha)+2\alpha}{16\pi
\left(r+2Br^3\right)^2}+
\frac{e^{-Ar^2}}{16\pi r^2}\left(Ar^2\left(5+Br^2+4\alpha
\right)\right.\\\nonumber
&-&\left.1-B^2r^4-2\alpha\right),\\\nonumber
\bar{P}_{r}&=&\frac{e^{-Ar^2}\left\{1+4Br^2(1+\alpha)+B^2r^4-A\left(r^2+B
r^4\right)+2\alpha\right\}-1-2\alpha}{16\pi r^2},\\\nonumber
\bar{P}_{t}&=&\frac{e^{-Ar^2}\left\{\left(1+Ar^2-Br^2\right)
\left(1+2Br^2\right)^2\right\}-1-3Br^2+2B^2r^4}{8\pi
r^2\left(1+2Br^2\right)}\\\nonumber
&+&\frac{e^{-Ar^2}\left\{1+4Br^2(1+\alpha)+B^2r^4-A\left(r^2+B
r^4\right)+2\alpha\right\}-1-2\alpha}{16\pi r^2},\\\nonumber
\bar{\Delta}&=&\frac{e^{-Ar^2}\left(\left(1+Ar^2-Br^2\right)
\left(1+2Br^2\right)^2\right)+\left(-1-3Br^2+2B^2r^4\right)}{8\pi
r^2\left(1+2Br^2\right)},\\\nonumber
q&=&\frac{1}{\sqrt{2}}\left[e^{-A r^2}r^2\left\{\left(1+B
r^2\right)\left(1+Br^2+2\alpha\right)-Ar^2
\left(3+Br^2+2\alpha\right)\right.\right.\\\nonumber
&+&\left.\left.e^{Ar^2}\left(1-2Ar^2\left(2+B
r^2\right)(1+\alpha)\right)\right\}\right]^{1/2}.
\end{eqnarray}
\begin{figure}\centering
\epsfig{file=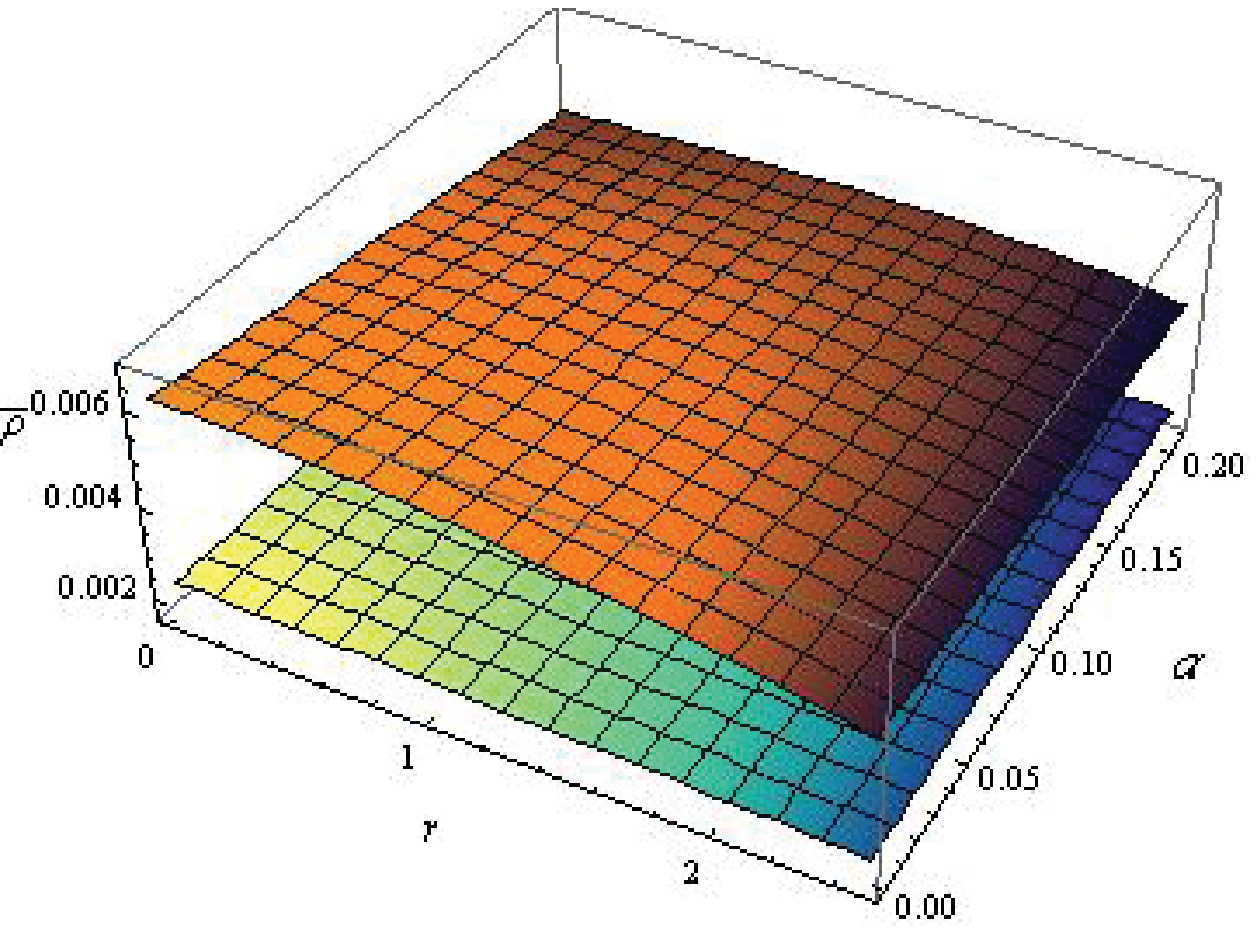,width=0.49\linewidth}
\epsfig{file=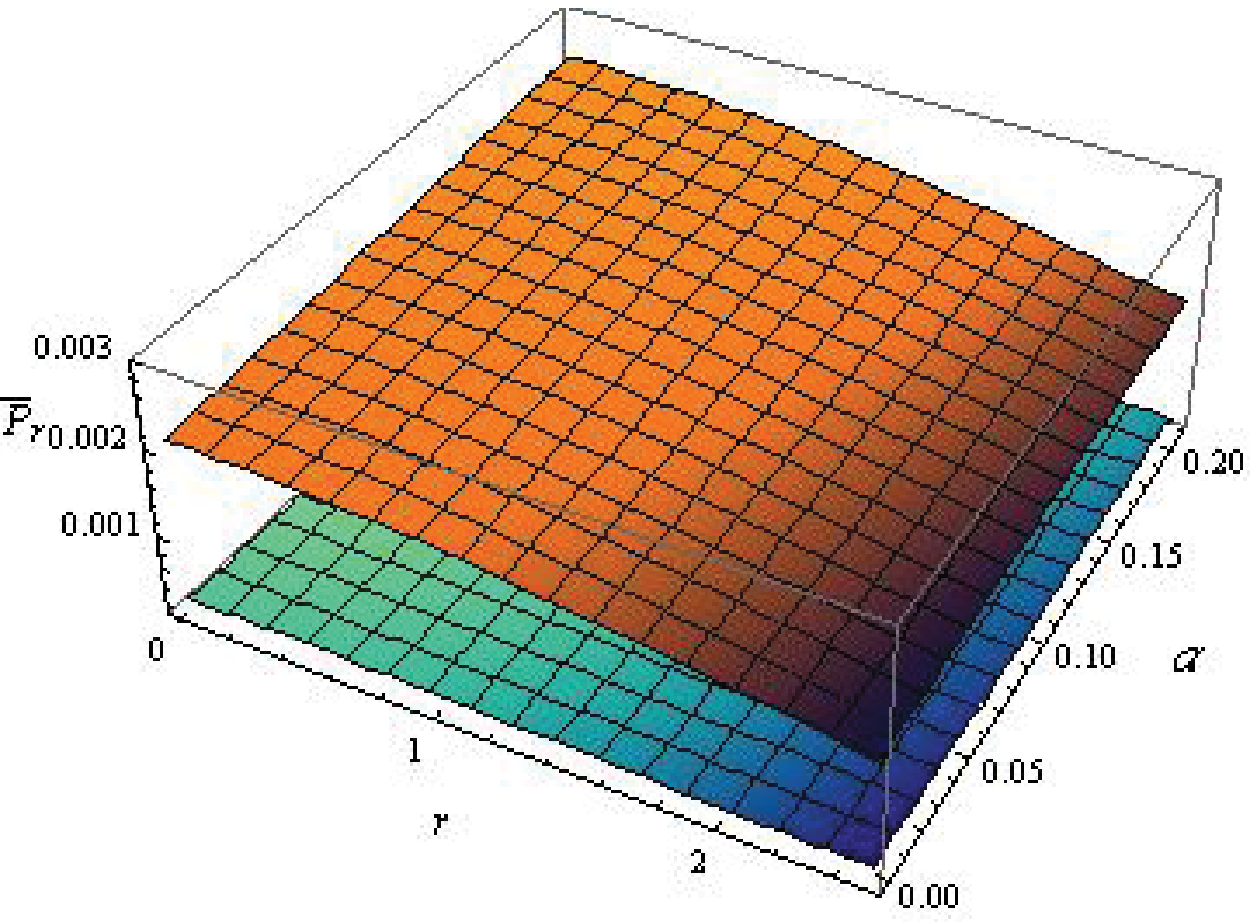,width=0.49\linewidth}
\epsfig{file=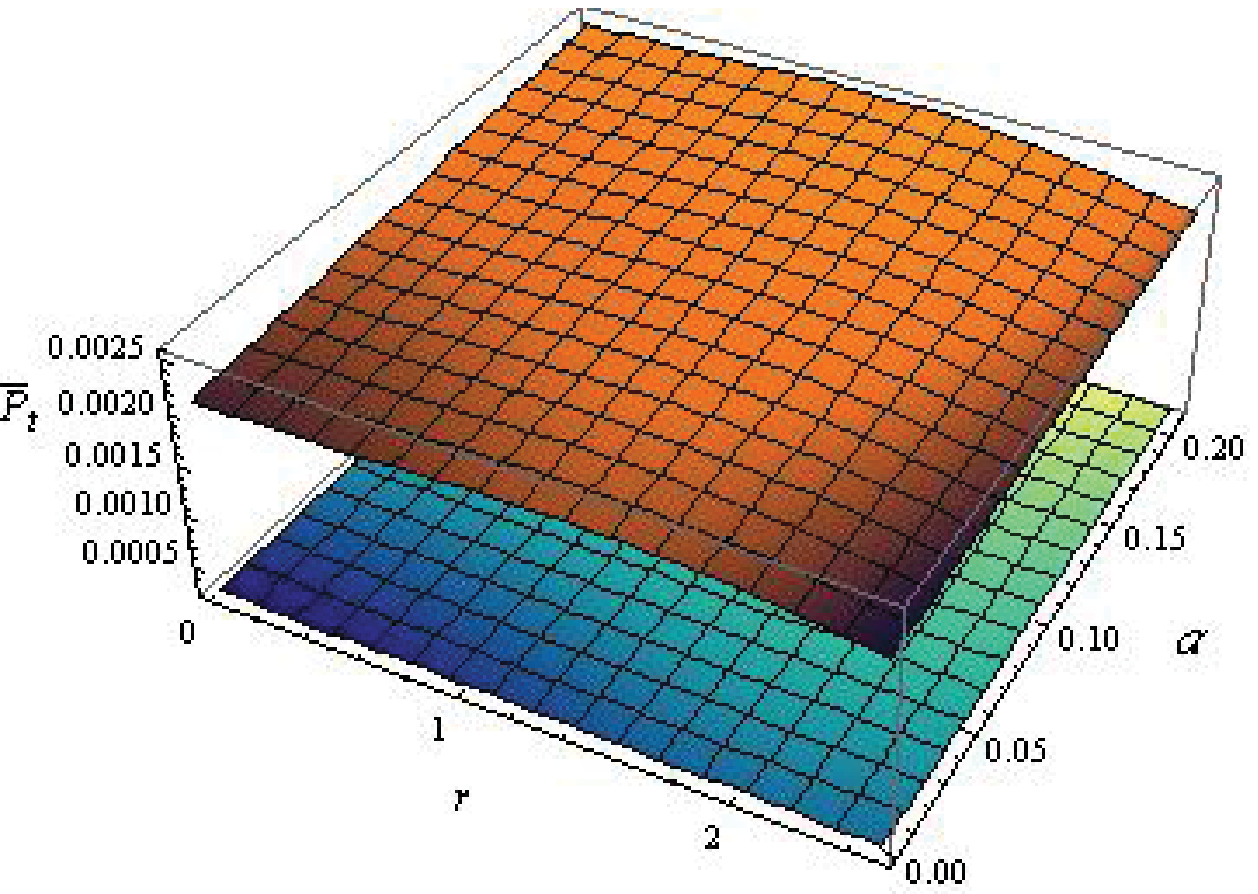,width=0.49\linewidth}
\epsfig{file=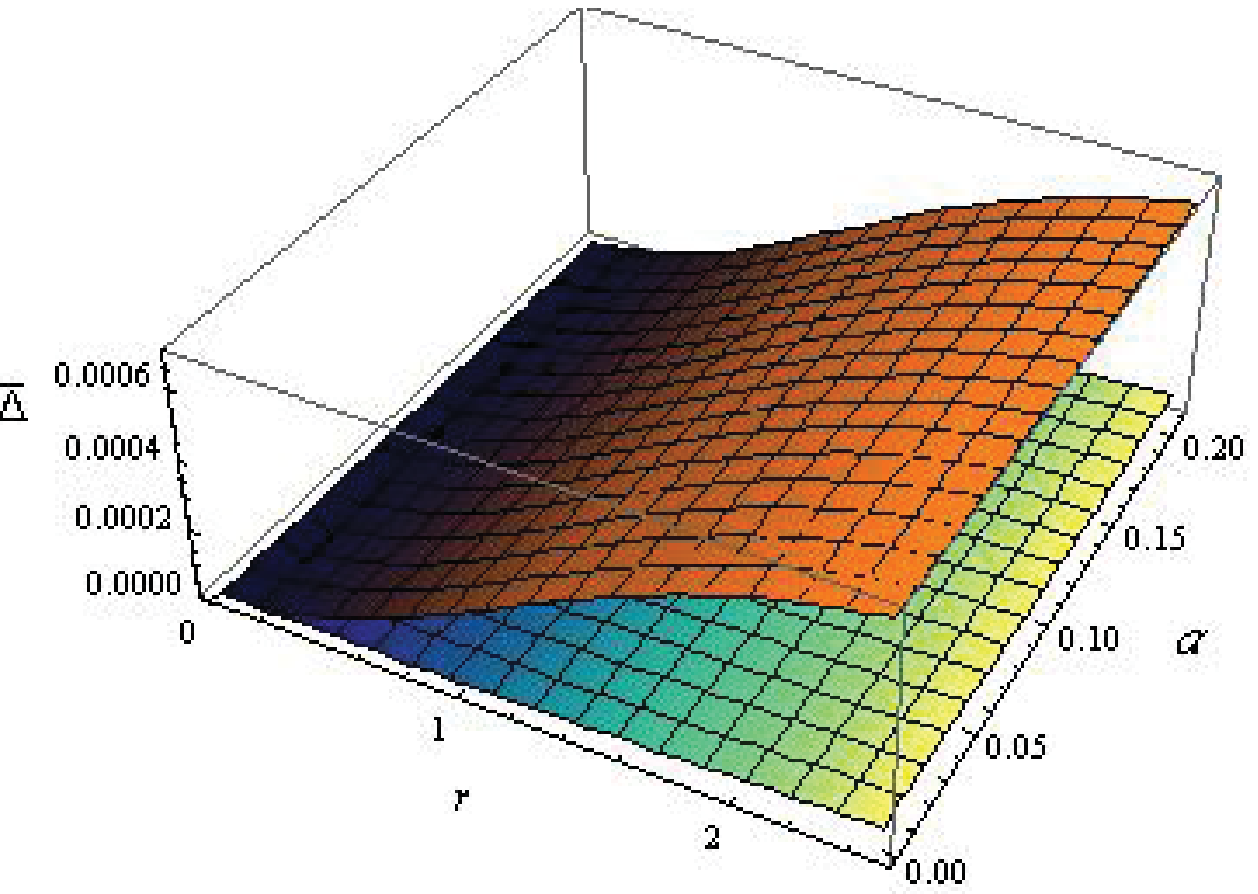,width=0.49\linewidth} \caption{Plots of
$\bar{\rho}$ (left plot, first row), $\bar{P}_{r}$ (right plot,
first row), $\bar{P}_{t}$ (left plot, second row) and $\bar{\Delta}$
(right plot, second row) versus $r$ and $\alpha$ with $Q_{0}=1$
(rust), $Q_{0}=3$ (multicolors), $M_{0}=1M_{\odot}$ and
$R=0.3M_{\odot}$ for solution \textbf{I}.}
\end{figure}

In order to examine physical characteristics of the above solution,
we plot this model. For graphical analysis, we fix the constant $A$
as given in Eq.(\ref{28}) while $B$ is a free parameter and will be
taken as mentioned in the isotropic case (Eq.(\ref{23})). The
compact stars demand that the behavior of energy density and radial
pressure should be positive, finite and maximum in the interior of
compact stars. The plot of $\bar{\rho}$ for two values of $Q_{0}$ is
shown in the left plot (first row of Figure \textbf{1}). We observe
that density is maximum in the interior and monotonically decreases
with increasing $r$. It is found that density attains larger values
for $Q_{0}=1$ while $Q_{0}=3$ yields smaller $\bar{\rho}$ leading to
the fact that increase in charge makes the sphere less dense.
Moreover, we find that $\bar{\rho}$ remains constant with increasing
$\alpha$.

The behavior of $\bar{P}_{r}$ is similar to that of density for
increasing $Q_{0}$, $r$ and $\alpha$ (right plot, first row of
Figure \textbf{1}). The plot of $\bar{P}_{t}$ (left plot, second row
of Figure \textbf{1}) shows that tangential pressure decreases with
increasing $r$ while corresponding to $\alpha$, it increases. It is
also found that the generic anisotropy remains same with increasing
coupling constant $\alpha$ while it acquires smaller values for
larger $Q_{0}$ (right plot, second row of Figure \textbf{1}).
\begin{figure}\centering
\epsfig{file=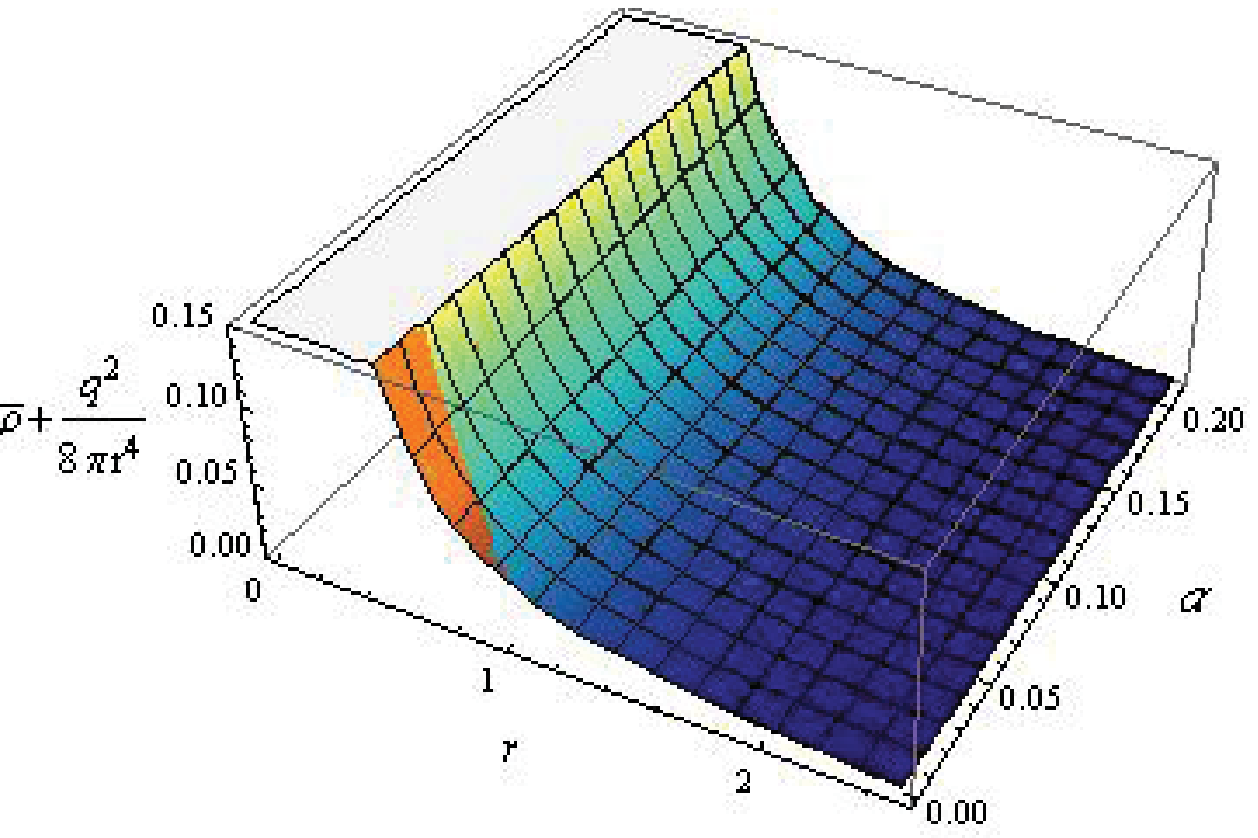,width=0.49\linewidth}
\epsfig{file=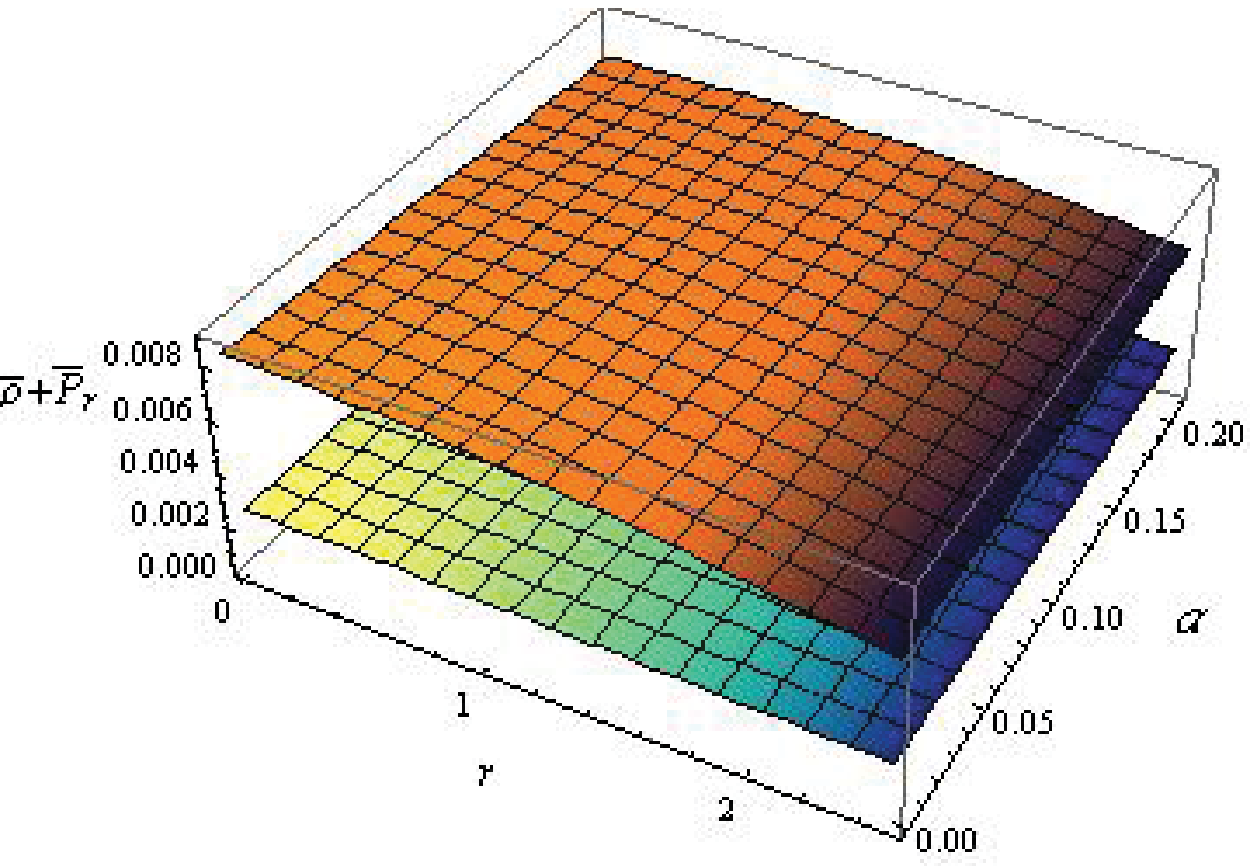,width=0.49\linewidth}
\epsfig{file=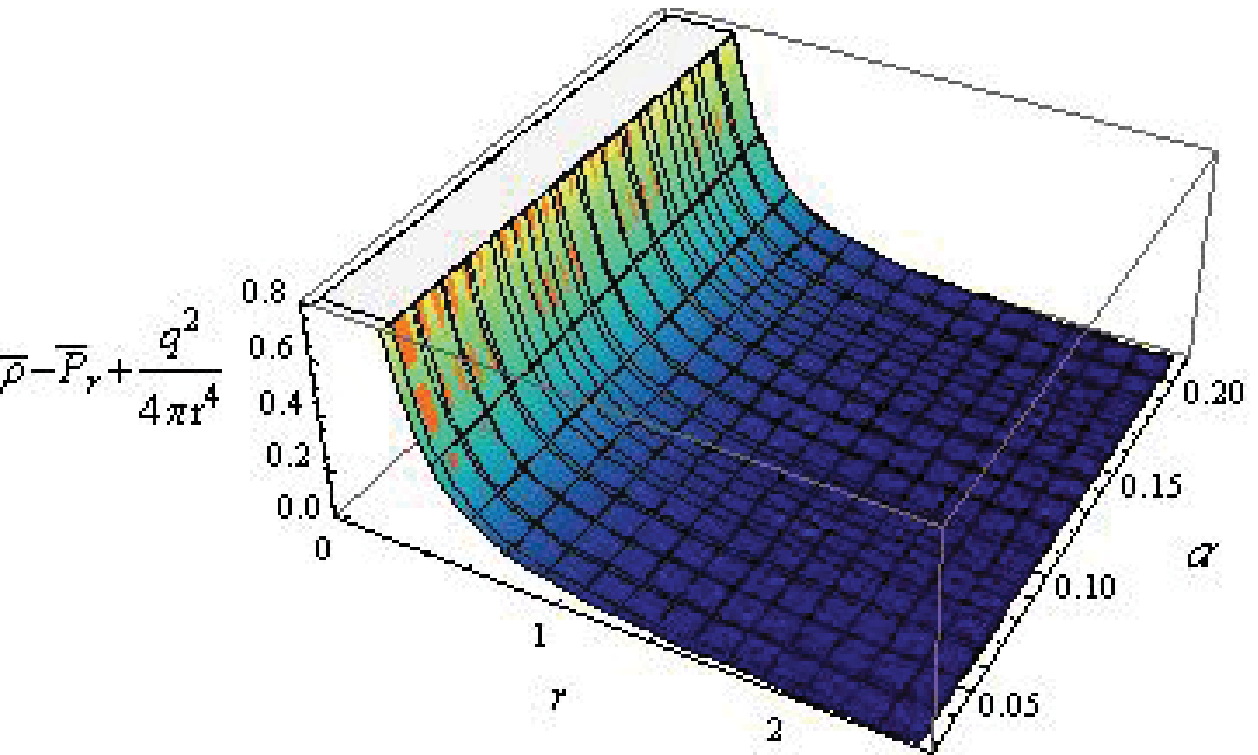,width=0.49\linewidth}
\epsfig{file=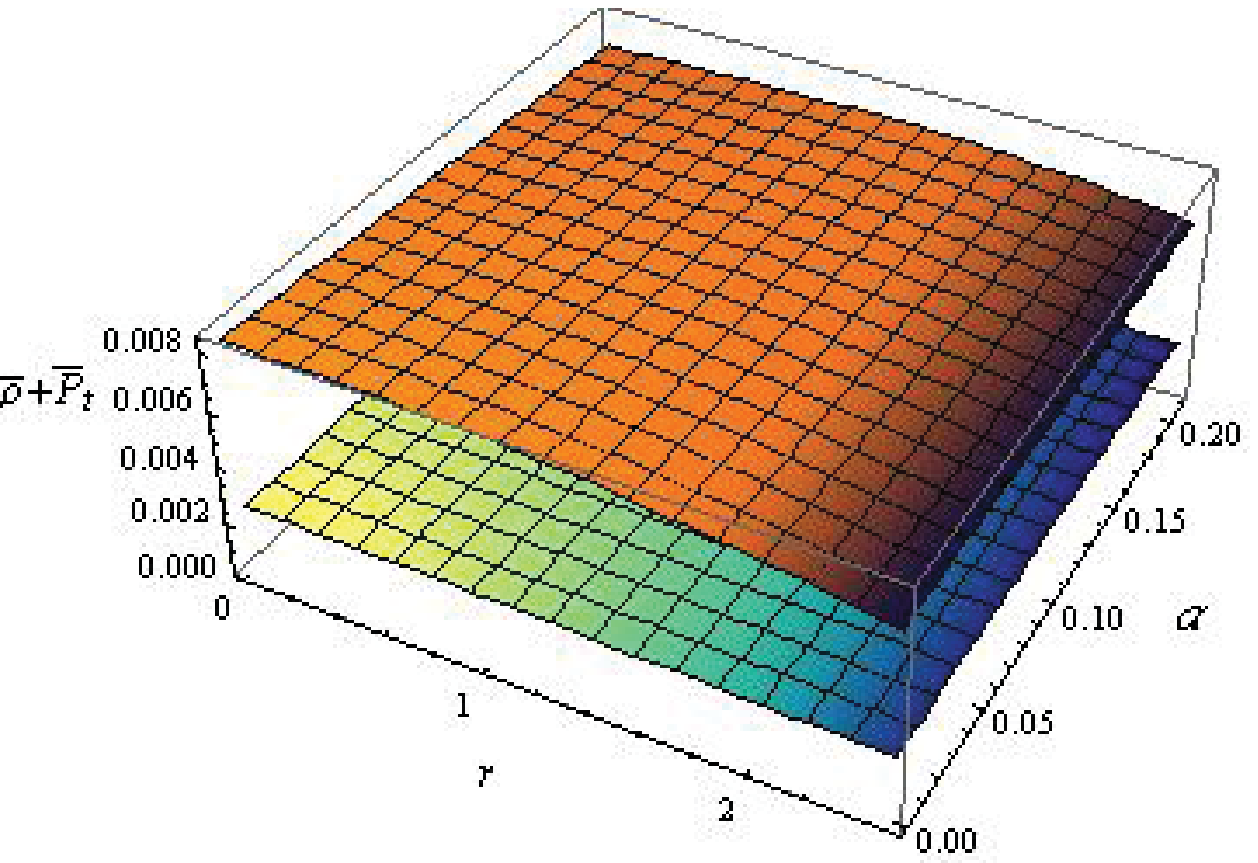,width=0.49\linewidth}
\epsfig{file=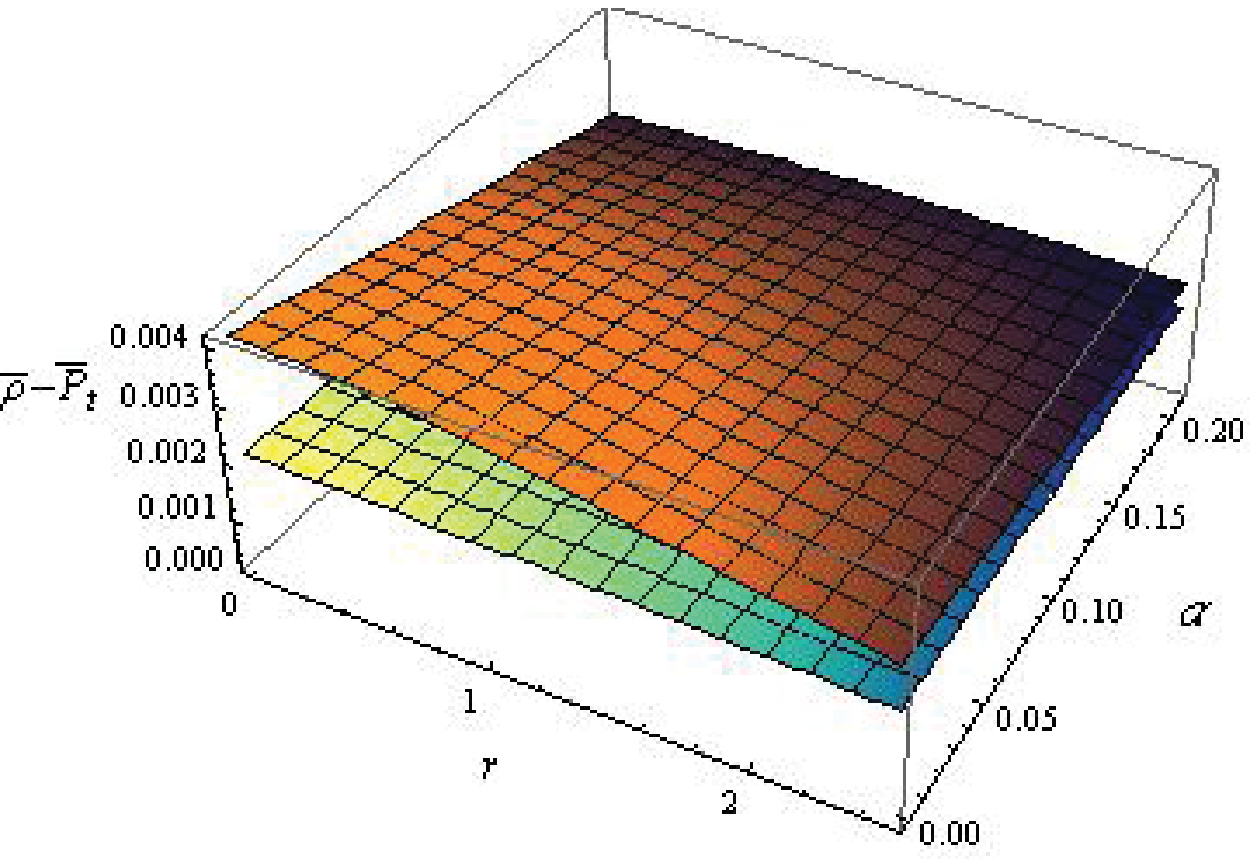,width=0.49\linewidth}
\epsfig{file=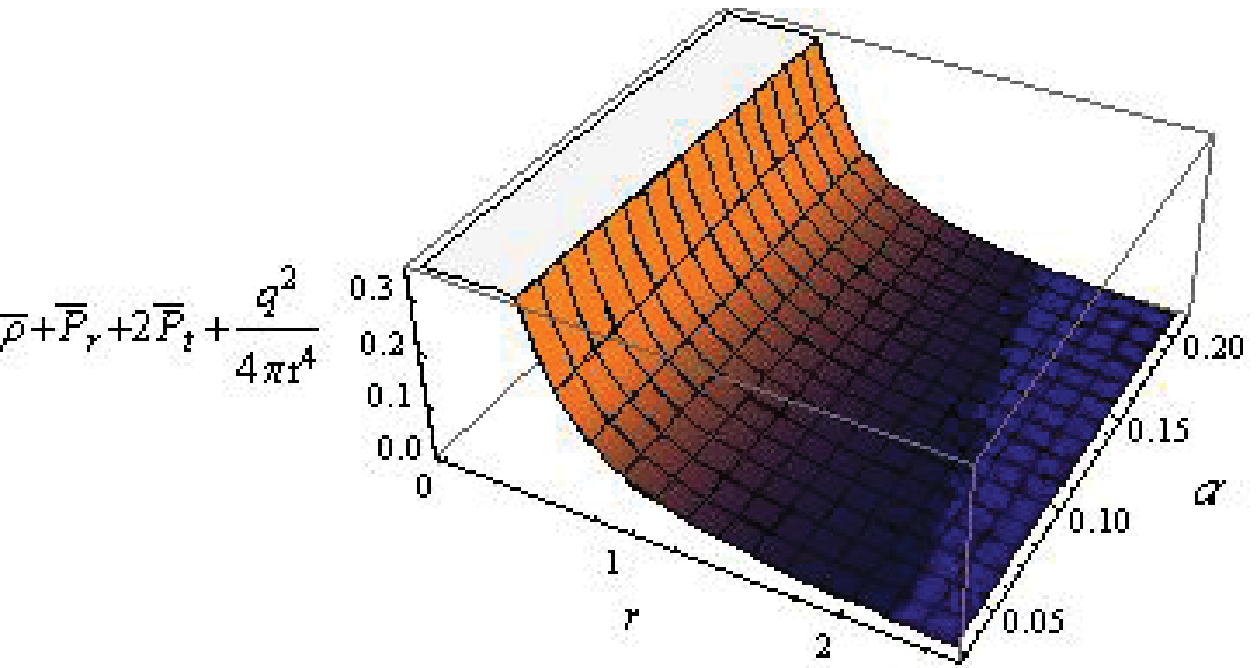,width=0.49\linewidth} \caption{Plots of energy
conditions versus $r$ and $\alpha$ with $Q_{0}=1$ (rust), $Q_{0}=3$
(multicolors), $M_{0}=1M_{\odot}$ and $R=0.3M_{\odot}$ for solution
\textbf{I}.}
\end{figure}
\begin{figure}\centering
\epsfig{file=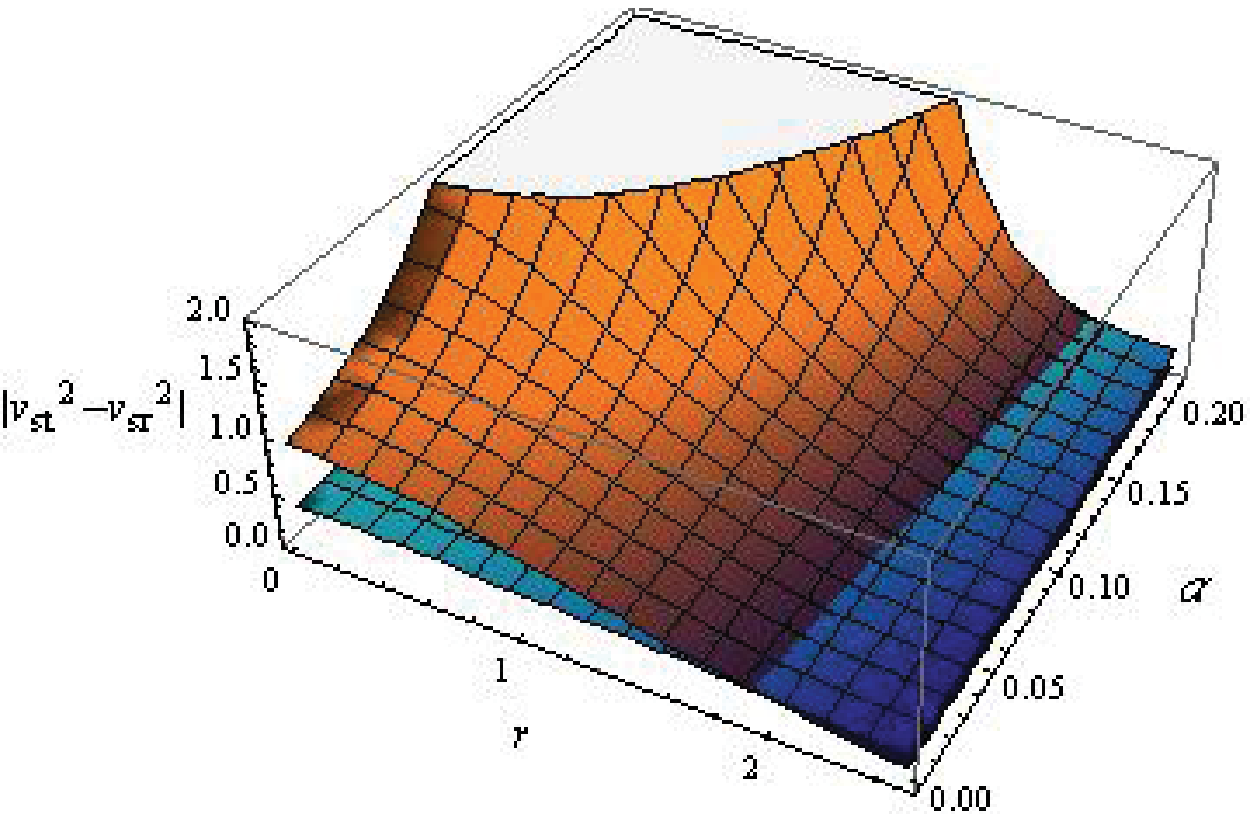,width=0.49\linewidth} \caption{Plot of
$|v_{st}^2-v_{sr}^2|$ versus $r$ and $\alpha$ with $Q_{0}=1$ (rust),
$Q_{0}=3$ (multicolors), $M_{0}=1M_{\odot}$ and $R=0.3M_{\odot}$ for
solution \textbf{I}.}
\end{figure}

In order to check physical viability of the resulting solutions, we
investigate the behavior of energy conditions which are the
constraints imposed on the energy-momentum tensor and describe
physically realistic matter distribution. For charged anisotropic
fluid configuration, these conditions turn out to be
\begin{eqnarray}\nonumber
&&\bar{\rho}+\frac{q^2}{8\pi r^4}\geq 0,\quad
\bar{\rho}+\bar{P}_{r}\geq 0,\quad \bar{\rho}+\bar{P}_{t}\geq
0,\quad \bar{\rho}-\bar{P}_{r}+\frac{q^2}{4\pi r^4}\geq
0,\\\nonumber &&\bar{\rho}-\bar{P}_{t}\geq 0,\quad
\bar{\rho}+\bar{P}_{r}+2\bar{P}_{t}+\frac{q^2}{4\pi r^4}\geq 0.
\end{eqnarray}
These are shown in Figure \textbf{2} which indicate that all the
conditions are satisfied confirming the physically viability of the
developed anisotropic solution. The stability is analyzed through
sound speed condition, i.e., $0<|v_{st}^2-v_{sr}^2|\leq1$. The plots
of stability condition for $Q_{0}=1,3$ are shown in Figure
\textbf{3}. It is found that $|v_{st}^2-v_{sr}^2|\leq1$ when
$Q_{0}=1$ for very small values of $\alpha$ while it is violated
with increasing $\alpha$. As the value of charge parameter is
increased, i.e., $Q_{0}=3$, stability criterion is fulfilled for all
values of $\alpha$ leading to the result that stability of charged
anisotropic sphere is enhanced with increasing charge parameter.

\subsection{Solution II}

In this case, we consider specific form of $\Theta_{0}^{0}$ to
obtain second type of anisotropic solution. The constraint is taken
as
\begin{equation}\nonumber
\Theta_{0}^{0}=\rho.
\end{equation}
Making use of Eqs.(\ref{12}) and (\ref{20}) in the above equation,
it follows that
\begin{equation}\nonumber
g^{*'}-\frac{g^{*}}{r}-8\pi
r\left\{\frac{e^{-Ar^2}}{16\pi}\left(5A-B(B-A)r^2-\frac{1}{r^2}\right)
+\frac{1}{16\pi
r^2}\right\}=0,
\end{equation}
whose solution is
\begin{eqnarray}\nonumber
g^{*}&=&rD+\frac{1}{8A^{3/2}}\left\{2e^{-Ar^2}\sqrt{A}\left(B^2r^2
+A(2-2e^{Ar^2}+Br^2)\right)\right.\\\label{31}
&+&\left.\left(14A^2+AB-B^2\right)\sqrt{\pi}r\text{Erf}(\sqrt{A}r)\right\},
\end{eqnarray}
where $``\text{Erf}"$ indicates the error function and $D$ is an
integration constant. By following the same procedure as for
solution \textbf{I}, we find the matching conditions as
\begin{eqnarray}\label{32}
&&1-e^{-Ar^2}(1+2BR^2)-\alpha(1+2BR^2)\left(RD+\frac{G}{8A^{3/2}}\right)
=0,\\\label{33}
&&\frac{2M_{0}}{R}+\frac{\mathcal{Q}^{2}-Q_{0}^{2}}{R^2}-\alpha
RD-\frac{\alpha
e^{-AR^2}G}{8A^{3/2}}=\frac{2\mathcal{M}}{R},\\\label{34}
&&1-\frac{2M_{0}}{R}+\frac{Q_{0}^{2}}{R^2}-e^{BR^2+C}+\alpha
RD+\frac{\alpha e^{-AR^2}G}{8A^{3/2}}=0,
\end{eqnarray}
where
\begin{eqnarray}\nonumber
G&=&2e^{-AR^2}\sqrt{A}\left(B^2R^2+A(2-2e^{AR^2}+BR^2)\right)
+\left(14A^2+AB-B^2\right)\\\nonumber
&\times&\sqrt{\pi}R\text{Erf}(\sqrt{A}R).
\end{eqnarray}
In this case, the anisotropic solution is obtained as
\begin{eqnarray}\nonumber
\bar{\rho}&=&\frac{e^{-Ar^2}\left(e^{Ar^2}-1-B^2r^4+Ar^2\left(5+B
r^2\right)\right)(1-\alpha )}{16\pi r^2},\\\nonumber
\bar{P}_{r}&=&\frac{1}{16\pi}\left\{\frac{1}{r^2}+\frac{e^{-Ar^2}}{r^2}
\left(A-4B-1+B(A-B)r^2\right)-\frac{2\alpha\left(2
Br^2+1\right)}{r^2}\right.\\\nonumber
&\times&\left.\left(\frac{e^{-Ar^2}}{4A}\left(B^2r^2+A
\left(2-Br^2-e^{Ar^2}(2-4rD)\right)\right)+\frac{\left(14
A^2+AB-B^2\right)}{8 A^{3/2}}\right.\right.\\\nonumber
&\times&\left.\left.\sqrt{\pi}r\text{Erf}\left(\sqrt{A}r\right)\right)
\right\},\\\nonumber
\bar{P}_{t}&=&\frac{e^{-A r^2}}{128A^{3/2}\pi r^2}\left[2\sqrt{A}
\left\{B^2r^2\left(3+3Br^2-2B^2r^4\right)\alpha
-2A^2r^2\left(1+Br^2\right)\right.\right.\\\nonumber
&\times&\left.\left.\left(2+\left(5+Br^2\right)\alpha\right)
-A\left(B^2r^4(5\alpha-4)-(8+4B^3r^6)
\alpha-Br^2(16+5\alpha)\right.\right.\right.\\\nonumber
&-&\left.\left.\left.4+4e^{Ar^2}\left(1+\alpha
\left(2-3rD+Br^2(2-3rD)-B^2r^4(1-2rD)\right)\right)\right)\right\}\right.\\\nonumber
&+&\left.\left(14A^2+AB-B^2\right)e^{Ar^2}\sqrt{\pi}r\left(3+3Br^2-2
B^2r^4\right)\alpha\text{Erf}\left(\sqrt{A}r\right)\right],\\\nonumber
\bar{\Delta}&=&\frac{-e^{-Ar^2}\left(1+Br^2\right)\alpha
}{128A^{3/2}\pi r^2}\left[2\sqrt{A}\left\{2A^2r^2\left(5+B
r^2\right)-B^2r^2\left(1-2Br^2\right)\right.\right.\\\nonumber
&-&\left.\left.A\left(4-5Br^2+4B^2r^4-e^{Ar^2}\left(4-4rD-4Br^2
(1-2rD)\right)\right)\right\}\right.\\\nonumber
&+&\left.\left(14A^2+AB-B^2\right)e^{Ar^2}
\sqrt{\pi}r\left(-1+2Br^2\right)\text{Erf}\left(\sqrt{A}r\right)\right],\\\nonumber
q&=&\frac{1}{2\sqrt{2}}\left[\frac{-e^{-Ar^2}
r^2}{A^{3/2}}\left(2\sqrt{A}\left(2A^2r^2\left(3+Br^2+2e^{Ar^2}\left(2+B
r^2\right)\right)+B^2r^2\alpha\right.\right.\right.\\\nonumber
&+&\left.\left.\left.A\left(e^{Ar^2}(-2+\alpha(4rD-2))-2B^2r^4-2(1-\alpha)-Br^2
(4+\alpha)\right)\right)\right.\right.\\\nonumber
&+&\left.\left.\left(14A^2+A
B-B^2\right)e^{Ar^2}\sqrt{\pi}r\alpha\text{Erf}\left(\sqrt{A}
r\right)\right)\right]^{1/2}.
\end{eqnarray}
\begin{figure}\centering
\epsfig{file=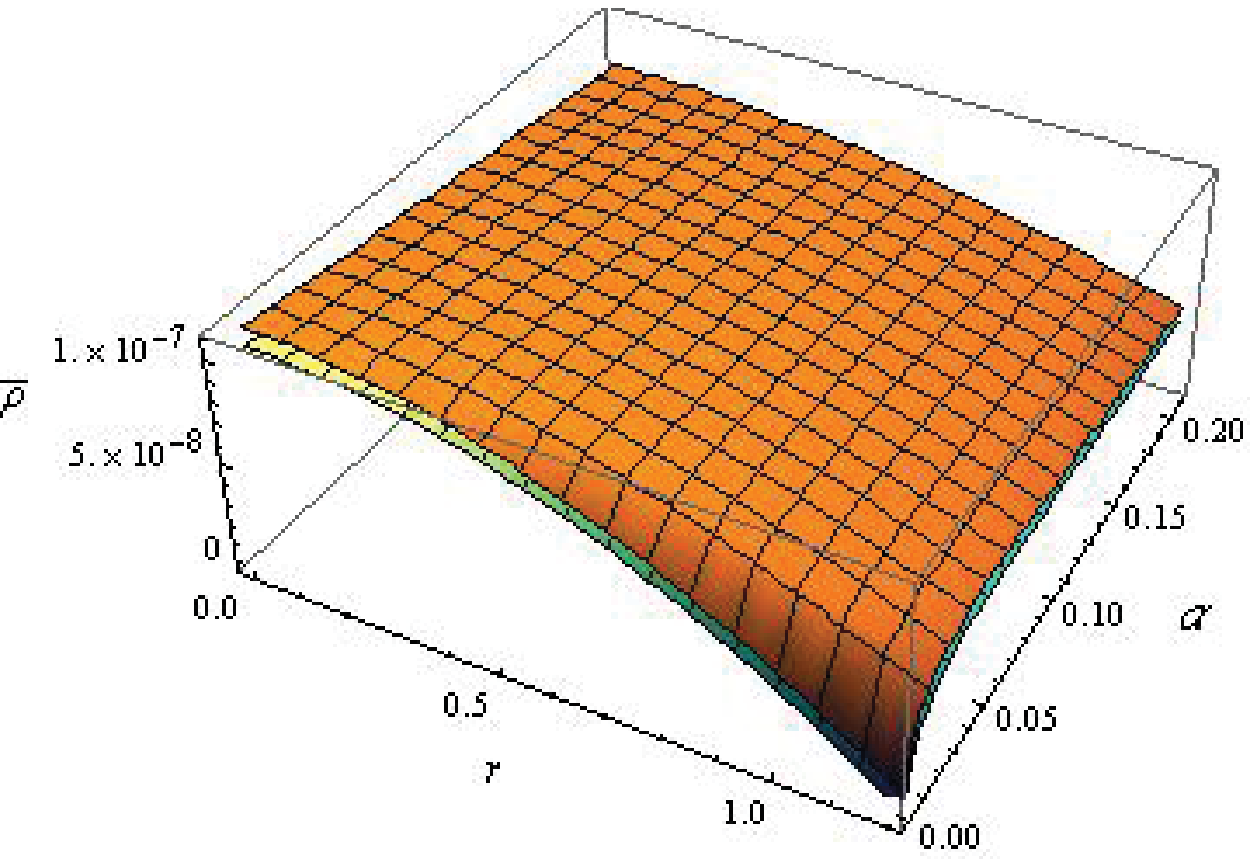,width=0.49\linewidth}
\epsfig{file=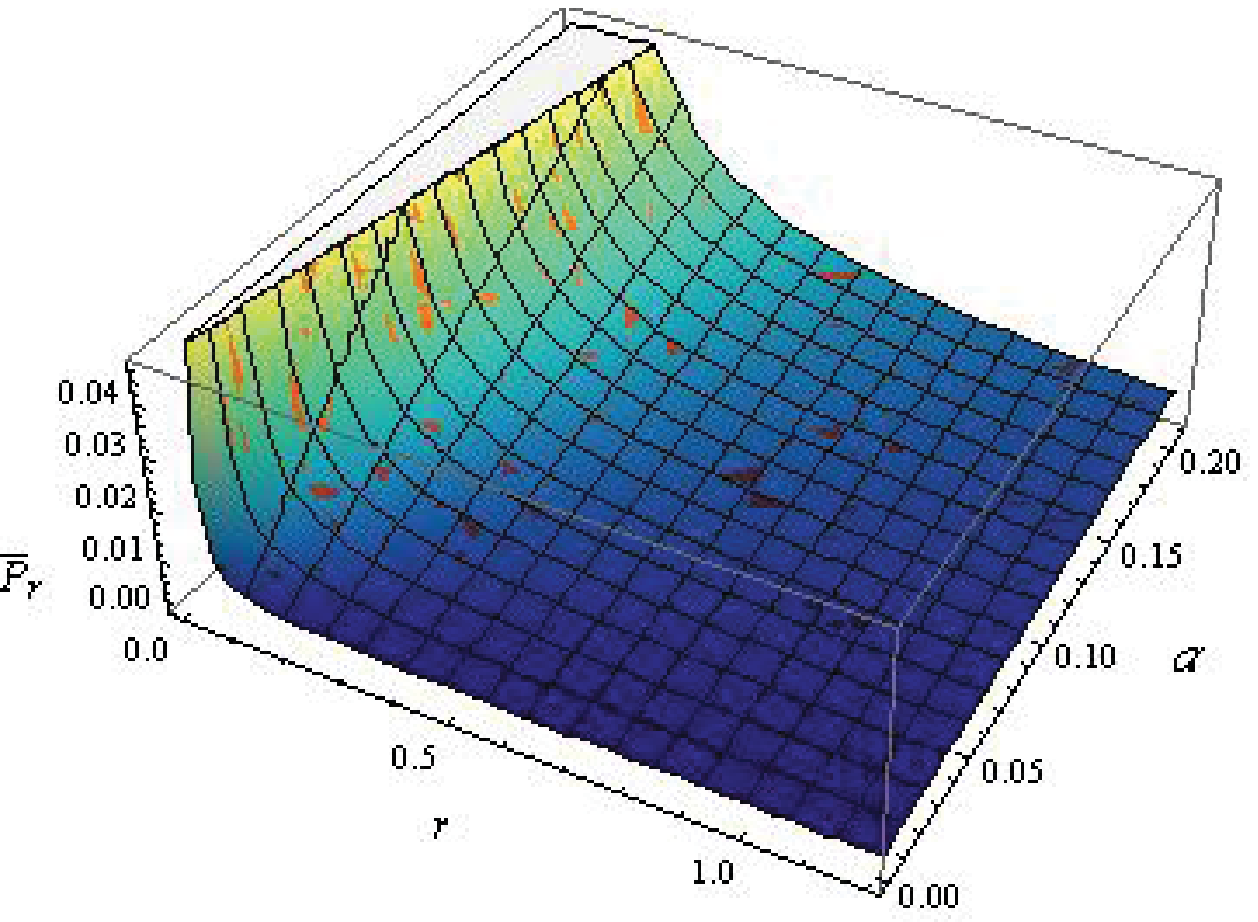,width=0.49\linewidth}
\epsfig{file=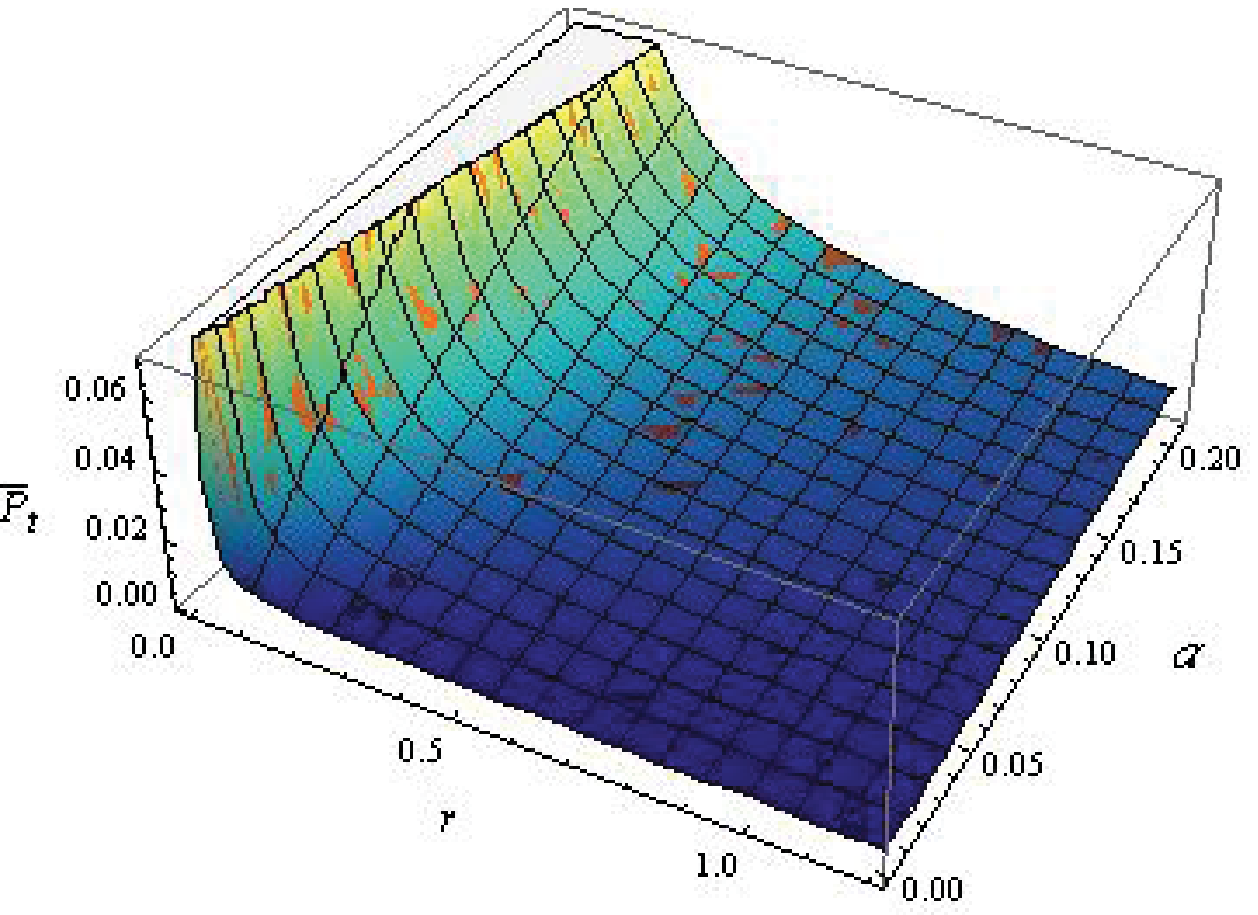,width=0.49\linewidth}
\epsfig{file=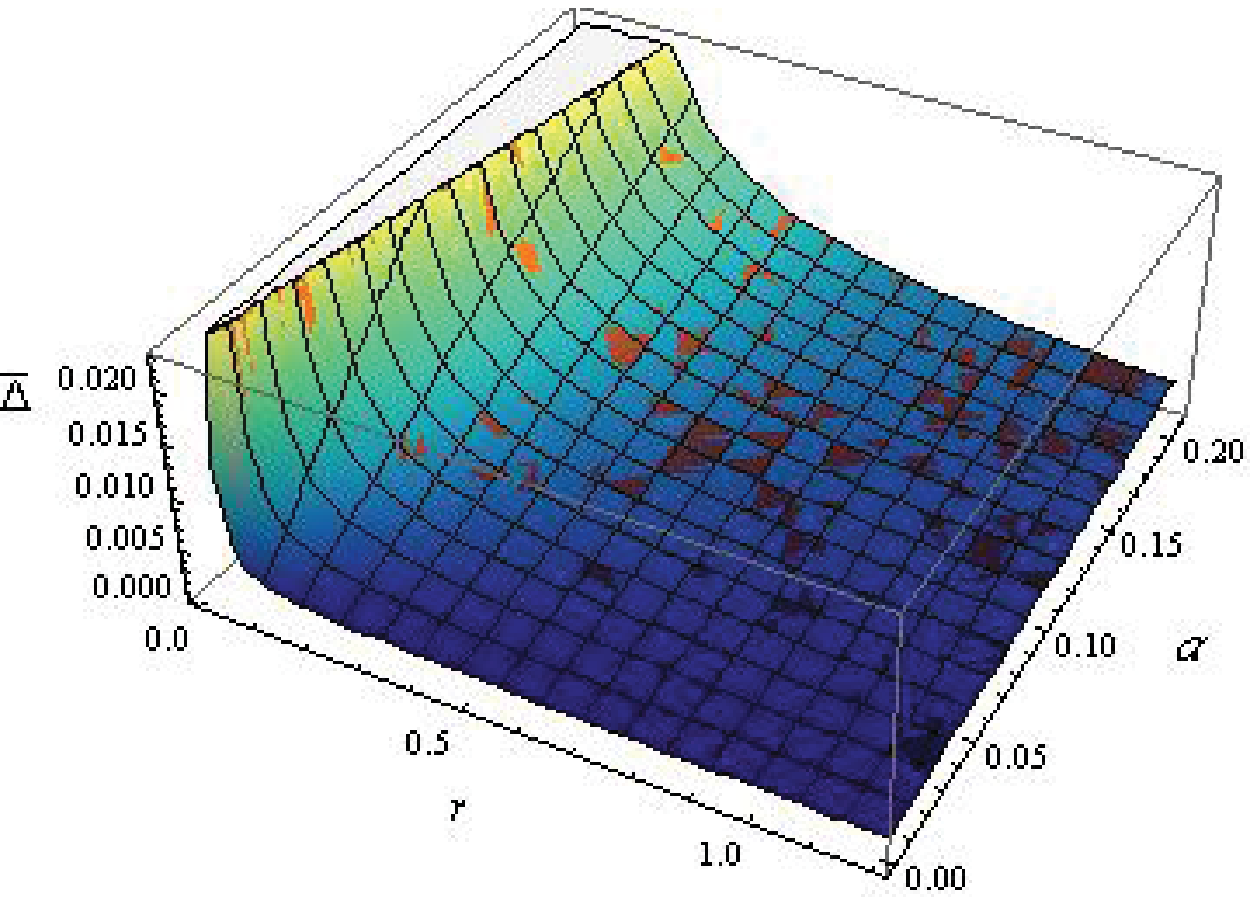,width=0.49\linewidth} \caption{Plots of
$\bar{\rho}$ (left plot, first row), $\bar{P}_{r}$ (right plot,
first row), $\bar{P}_{t}$ (left plot, second row) and $\bar{\Delta}$
(right plot, second row) versus $r$ and $\alpha$ with $Q_{0}=1$
(rust), $Q_{0}=6$ (multicolors), $M_{0}=1M_{\odot}$ and
$R=0.01M_{\odot}$ for solution \textbf{II}.}
\end{figure}

In order to plot the developed solution, we fix the constant $B$ by
solving Eqs.(\ref{32}) and (\ref{34}) (which is not mentioned here
due to lengthy expression) while $A$ is a free parameter which will
be taken as given in Eq.(\ref{23}) and $D=1$. The behavior of
density and radial/tangential pressure (Figure \textbf{4})
corresponding to the variation in $r$ is similar to that obtained in
solution \textbf{I}. However, the behavior of $\bar{\rho}$ and
$\bar{P}_{r}$ is different with respect to $\alpha$, i.e., it is an
increasing function as the parameter $\alpha$ is increased while the
behavior of $\bar{P}_{t}$ is consistent with solution \textbf{I}.
This shows that $\Theta_{\alpha\beta}$ increases the compactness of
spherical matter configuration. Moreover, we find that the change in
charge parameter does not yield much difference between the values
of all physical parameters. We find that the generated anisotropy is
greater for the larger values of $\alpha$ (last plot, Figure
\textbf{4}) and decreases towards surface which is opposite to the
anisotropic behavior in the absence of electromagnetic field.
\begin{figure}\centering
\epsfig{file=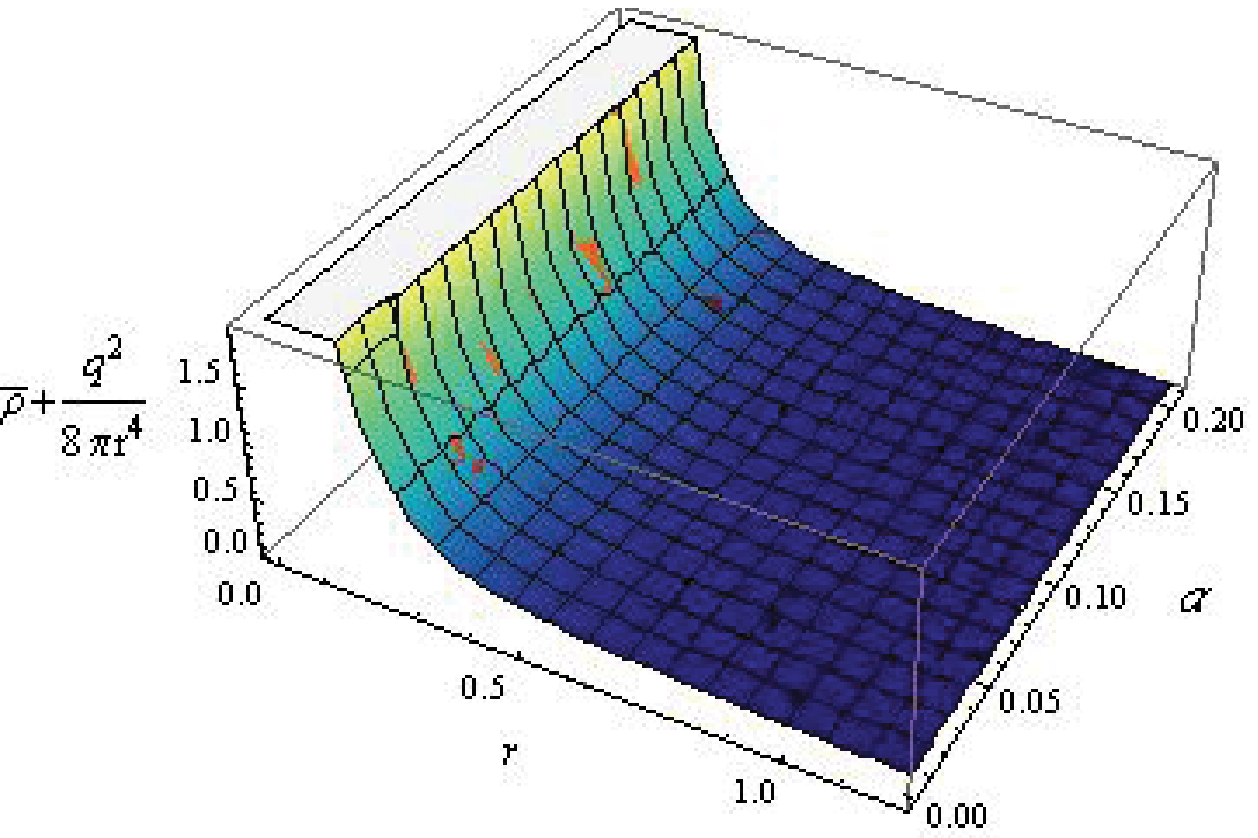,width=0.49\linewidth}
\epsfig{file=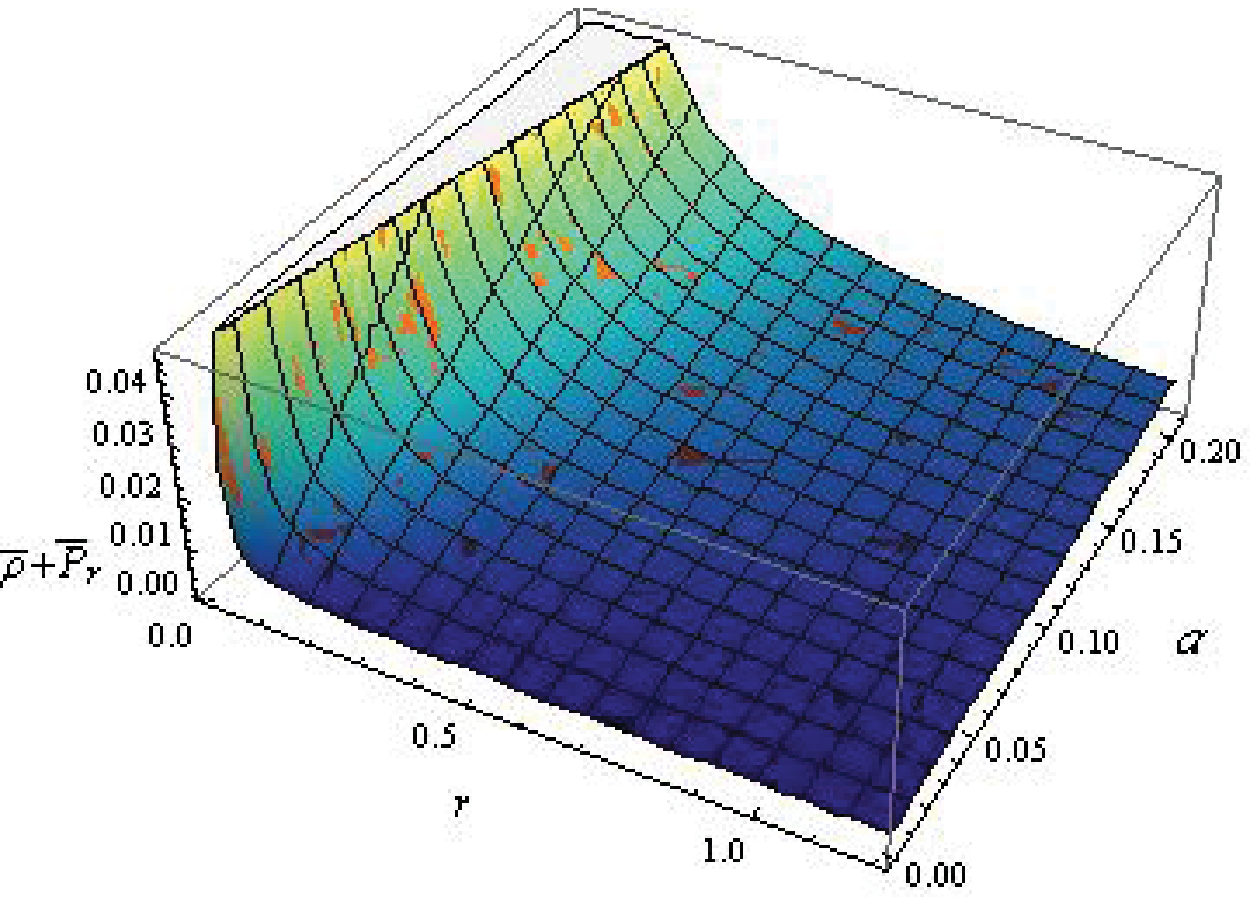,width=0.49\linewidth}
\epsfig{file=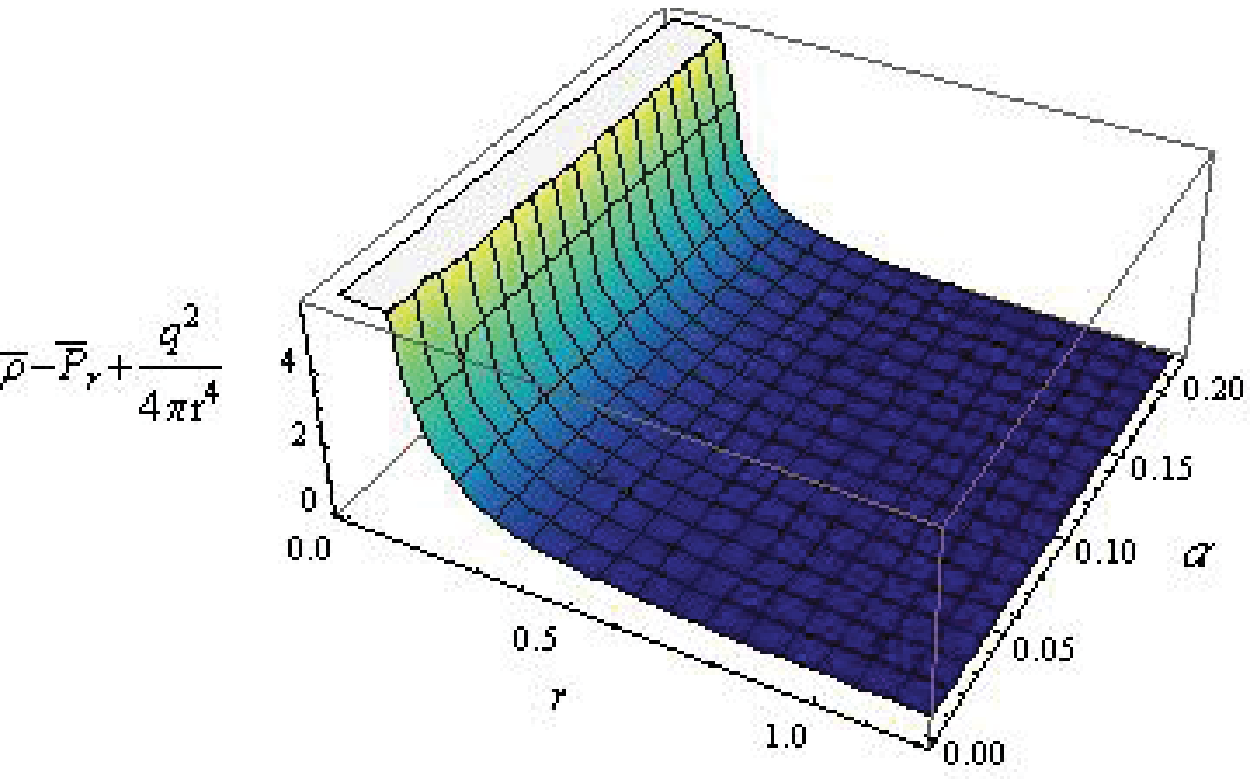,width=0.49\linewidth}
\epsfig{file=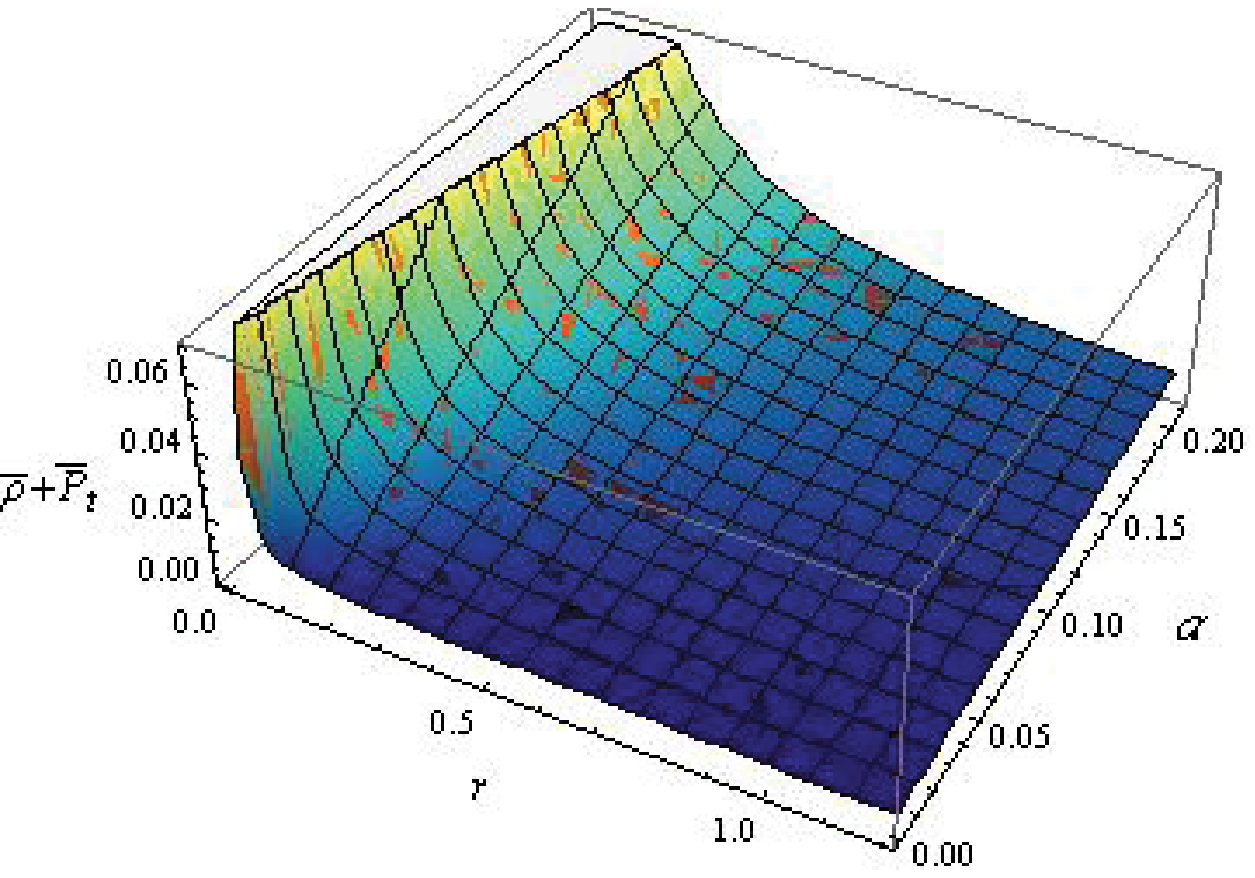,width=0.49\linewidth}
\epsfig{file=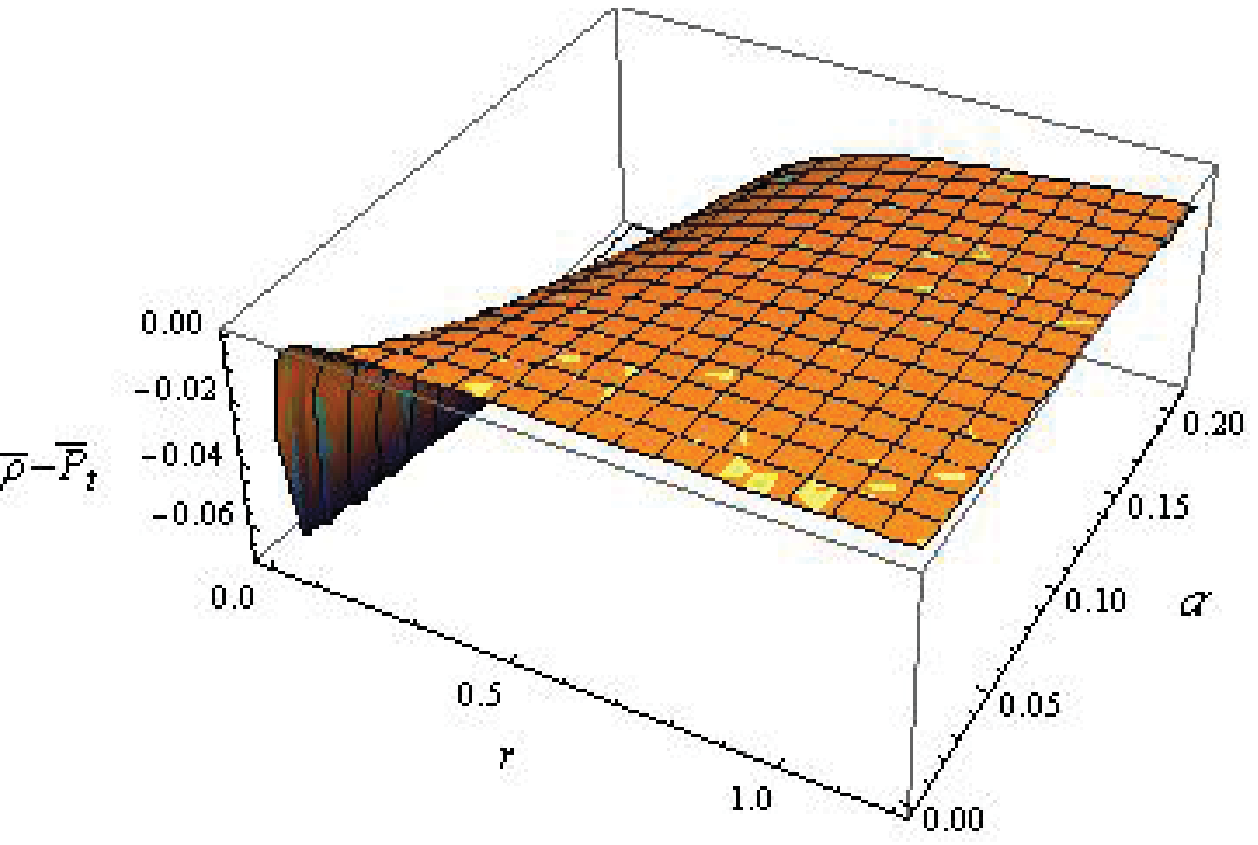,width=0.49\linewidth}
\epsfig{file=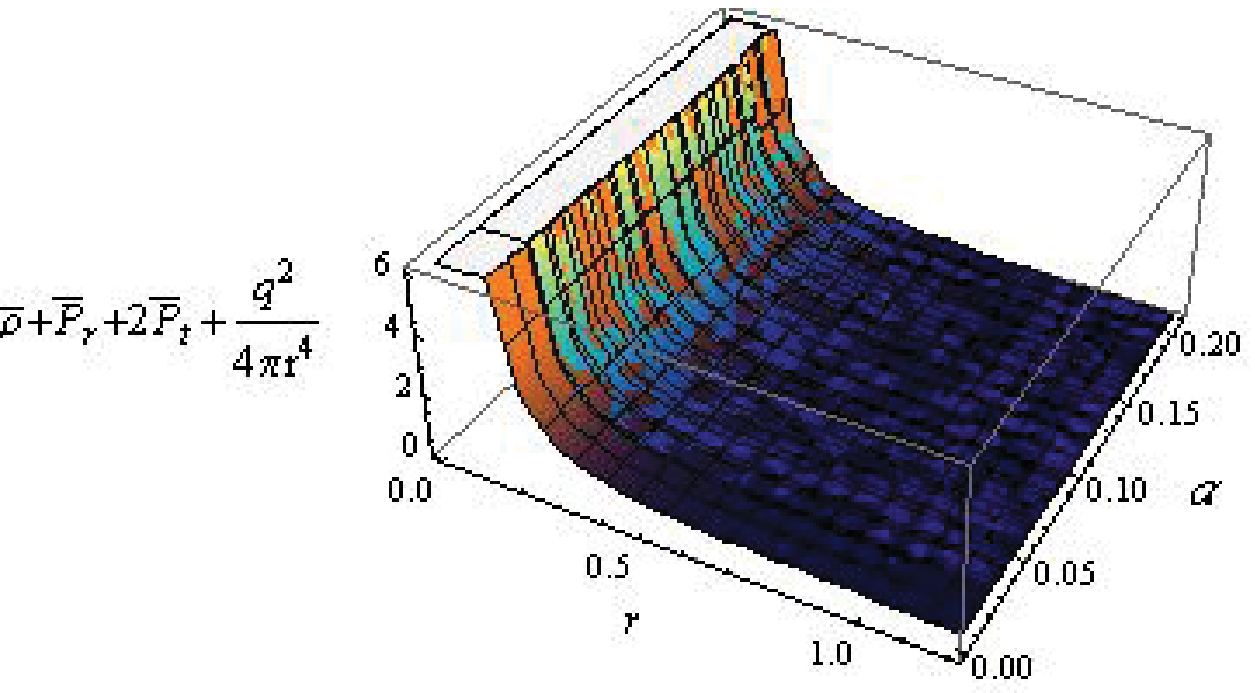,width=0.49\linewidth} \caption{Plots of energy
conditions versus $r$ and $\alpha$ with $Q_{0}=1$ (rust), $Q_{0}=6$
(multicolors), $M_{0}=1M_{\odot}$ and $R=0.01M_{\odot}$ for solution
\textbf{II}.}
\end{figure}
\begin{figure}\centering
\epsfig{file=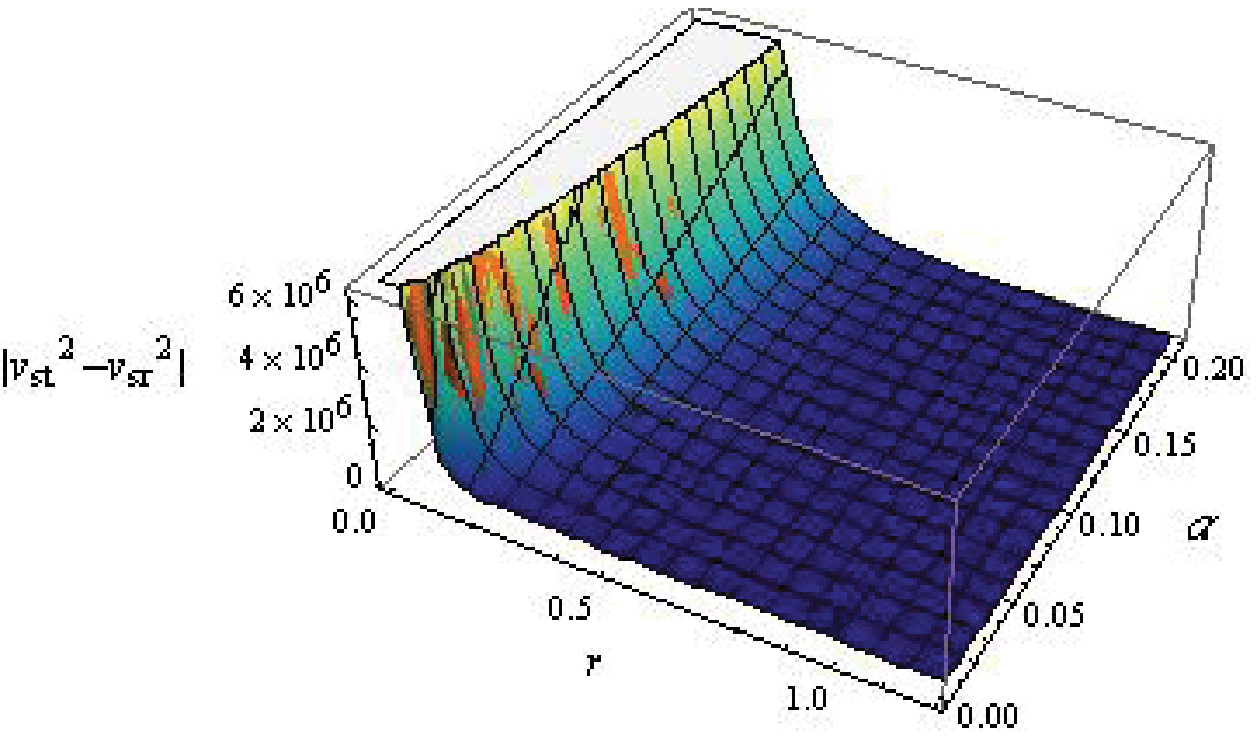,width=0.49\linewidth} \caption{Plot of
$|v_{st}^2-v_{sr}^2|$ versus $r$ and $\alpha$ with $Q_{0}=1$ (rust),
$Q_{0}=6$ (multicolors), $M_{0}=1M_{\odot}$ and $R=0.01M_{\odot}$
for solution \textbf{II}.}
\end{figure}

The plots of all energy conditions are shown in Figure \textbf{5}.
It is found that the resulting solution meets all the energy bounds
except $\bar{\rho}-\bar{P}_{t}$. This shows that the solution
\textbf{II} is not physically viable for both values of $Q_{0}$.
Furthermore, we plot the stability condition
$0<|v_{st}^2-v_{sr}^2|\leq1$ (Figure \textbf{6}) and obtain that it
is violated throughout the system.

\section{Final Remarks}

The search for interior solutions describing self-gravitating
systems has captivated the attention of many researchers. Recently,
the minimal gravitational decoupling technique has widely been used
to find exact solutions for interior constitution of stellar
objects. In this paper, we have explored exact solutions of the
charged anisotropic field equations from known isotropic model using
MGD approach. For this purpose, a new source is added to the charged
isotropic energy-momentum tensor which leads to the effective field
equations with anisotropic matter distribution. Then, we have
introduced a geometric deformation for the radial metric function of
the line-element (used in the known solution). This deformation
leads to two sets of the field equations: the first set is similar
to the standard Einstein equations for charged isotropic source
while the second one corresponds to the additional source and the
deformed metric coefficient. We have also formulated junction
conditions for the smooth matching of the interior region with the
exterior one described by the deformed Riessner-Nordstr\"{o}m
spacetime.

In order to seek anisotropic solutions, we have firstly considered
the known isotropic solution with electromagnetic field and then
incorporated the effects of source added to charged perfect fluid.
For this purpose, we have imposed two constraints depending upon
pressure and density leading to solutions \textbf{I} and
\textbf{II}, respectively. We have analyzed physical characteristics
of constructed models and found that density and radial/tangential
pressure exhibit viable behavior. The physical acceptability has
also been investigated through energy conditions. It is found that
the first solution fulfils these conditions while one of them is
violated for the solution \textbf{II}. We have examined the
stability through sound speed criterion and concluded that the first
model is stable whereas the second does not meet the stability
condition. Moreover, we have found that the increase in charge
parameter increases the stability of the first model. It is
interesting to mention here that the solution \textbf{I} is
physically acceptable as it satisfies all the conditions required
for stellar objects. We would like to point out here that such
conditions are not checked for the uncharged solutions \cite{8}.

\vspace{0.5cm}

\section*{Acknowledgement}

We would like to thank the Higher Education Commission, Islamabad,
Pakistan for its financial support through the \emph{Indigenous
Ph.D. Fellowship, Phase-II, Batch-III}.

\end{document}